\documentstyle[epsf,psfig]{mn_latex}

\newcommand{\mnref}[1]{\hangindent=0.5in \hangafter=1 #1 \par}
\newenvironment{refs}{\parindent=0pt}{\parindent=1.5em}

\def\etal{\hbox{\it et al.}$\,$}
\def\simlt{\lower.5ex\hbox{$\; \buildrel < \over \sim \;$}}
\def\simgt{\lower.5ex\hbox{$\; \buildrel > \over \sim \;$}}
\def\sext{{\sl SExtractor}}
\def\taurus2{{\sc Taurus-2}}
\def\lineunits{erg\,s$^{-1}$\,cm$^{-2}$}

\def\ergs2band{erg\,s$^{-1}$\,cm$^{-2}$\,band$^{-1}$}

\def\as{$''$}

\def\um{$\mu$m}

\def\MpcPer3{Mpc$^{-3}$}
\def\Mpc3{Mpc$^{3}$}
\def\eADU{e$^{-}\,$ADU$^{-1}$}
\def\ergphot{erg\,$\gamma^{-1}$}

\def\Oiii{{[{{\sc O}\,{\sc iii}}]~$\lambda\lambda$4959,5007}\/}
 
\def\Hb{{H$\beta$~$\lambda$4861}\/}

\begin{document}
\label{firstpage}

\title
[Detection and Measurement from Narrowband Scans]
{Detection and Measurement from Narrowband Tunable Filter Scans}
\author
[D.~H.~Jones,  P.~L.~Shopbell and J.~Bland-Hawthorn]
{D.~Heath~Jones$^1$, Patrick~L.~Shopbell$^2$ and Joss~Bland-Hawthorn$^3$ 
\vspace{0.5cm} \\ 
$^1$ European Southern Observatory, Casilla 19001, Santiago 19, Chile --- {\tt hjones@eso.org} \\ 
$^2$ Caltech, MC 105-24, Pasadena, CA 91125, USA --- {\tt pls@astro.caltech.edu} \\ 
$^3$ Anglo-Australian Observatory, PO Box 296, Epping, NSW 2121, Australia 
--- {\tt jbh@aaoepp.aao.gov.au} }

\maketitle

\begin{abstract}
The past five years have seen a rapid rise in the use
of tunable filters in many diverse fields of astronomy,
through Taurus Tunable Filter (TTF) instruments at the
Anglo-Australian and William Herschel Telescopes.
Over this time we have continually refined aspects 
of operation and developed a collection of special techniques 
to handle the data produced by these novel 
imaging instruments. In this paper, we review calibration
procedures and summarize the theoretical basis for Fabry-Perot 
photometry that is central to effective tunable imaging.
Specific mention is made of object detection and classification
from deep narrowband surveys containing several hundred objects
per field. We also discuss methods for recognizing 
and dealing with artefacts (scattered light,
atmospheric effects, etc.) which can seriously compromise the 
photometric integrity of the data if left untreated.
Attention is paid to the different families of
ghost reflections encountered, and strategies to minimise their
presence. 
In our closing remarks, future directions for tunable imaging 
are outlined and contrasted with the Fabry-Perot technology
employed in the current generation of tunable imagers.
\end{abstract}


\section{Introduction}
Imaging Fabry-Perot interferometers are now in common use at several 
major observatories and operate at both optical and infrared wavelengths. 
Traditionally, Fabry-Perots are employed to perform studies
of extended gaseous nebulae. Examples include outflow sources (starburst
and active galaxies, Herbig-Haro systems) and normal disk galaxies. 
Some groups have utilised the angular spectral coverage to detect diffuse
sources (HI/Ly$\alpha$ clouds in optical emission), by azimuthally
summing deep Fabry-Perot images (Haffner, Reynolds \& Tufte~1999). 
More recently, scanning Fabry-Perots have been used to construct spectral line 
profiles at many pixel positions across a large format CCD. In many 
instances, these spectra are used simply to obtain kinematic information 
(line widths, radial velocities) in a strong emission line 
({\em e.g.}~Laval \etal\ 1987; Cecil~1988; 
Veilleux, Bland-Hawthorn \& Cecil~1997; Shopbell \& Bland-Hawthorn~1998). 
At lower spectral resolutions, {\sl tunable filters} are a special
class of Fabry-Perot, with which a series of narrowband images, stepped in
wavelength, can be obtained (Atherton \& Reay~1981; 
Bland-Hawthorn \& Jones~1998a,b; Jones \& Bland-Hawthorn 2001).
Alternative modes see the Fabry-Perot used for time-series imaging
(Deutsch, Margon \& Bland-Hawthorn 1998; Tinney \& Tolley 1999), 
which we do not discuss here.
It is less common to see these instruments employed as 
spectro{\sl photometers} where the observed signal is calibrated to 
exoatmospheric flux units. This may be due, in part, 
to a perception that the nature of the Airy function makes
Fabry-Perots unreliable photometers.

In this review, we describe reliable methods for flux-calibrating 
Fabry-Perot data with the aid of worked examples.
The case of tunable imaging is special, in that photometric measurements
centre upon the measurement of low resolution spectra and extraction of
objects with candidate emission or absorption features. 
In this sense it is
analogous to multi-slit or multi-fibre spectroscopy.
However, because the tunable filter images {\sl all}
objects in the field at once, every object is a potential
candidate. Therefore, data reduction and methods need to be 
devised so as to be fully automated throughout.
In deep extragalactic surveys at high galactic latitude,
the number of objects can vary anywhere between 
a few hundred in high galactic latitude fields
to over a thousand in galaxy clusters.

This paper reviews all stages of analysis, discussing different
approaches where they exist.
Section 2 describes features of tunable filter use and Section 3 
defines terminology used throughout the rest of the paper.
Section 4 discusses different methods for the removal of instrumental
signatures while Section 5 reviews strategies for the detection
and classification of the many objects in a single field.
Flux calibration issues are discussed in Section 6 and concluding
remarks follow.
Many of the procedures described here have been written into 
software (Jones~1999).
Interested prospective users are encouraged to contact the authors
and visit the Taurus Tunable Filter (TTF) web page.\footnote{
The Taurus Tunable Filter web site is maintained at the
Anglo-Australian Observatory: http://www.aao.gov.au/ttf/
}


\section{The Nature of the Observations}

\begin{figure*} 
\psfig{file=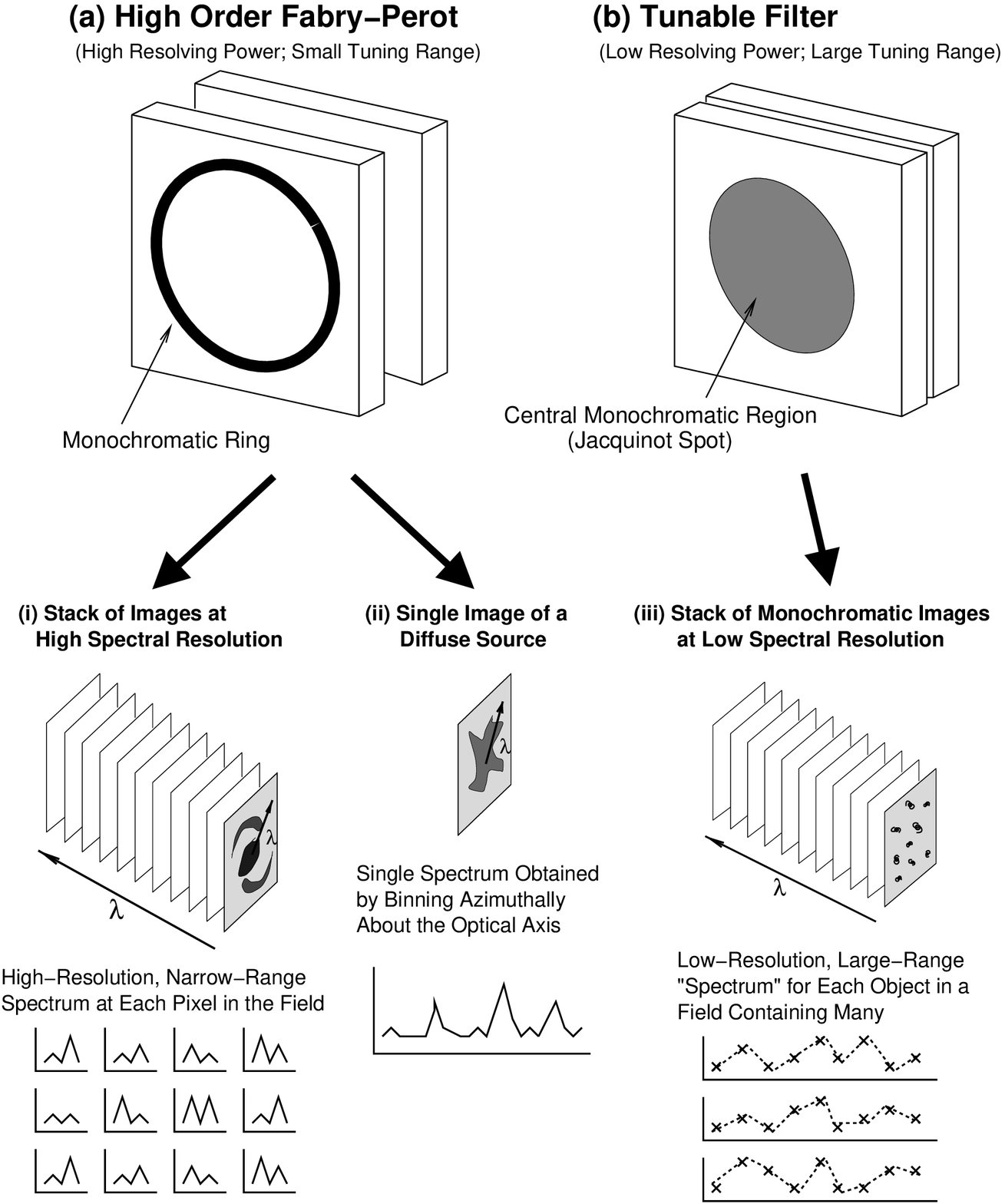,width=\hsize}
\caption[Different Modes of Fabry-Perot use]
{Different modes of Fabry-Perot use in the case of both ($a$)
high resolution instruments, and ($b$) tunable filters.}
\label{differentModes}
\end{figure*}

At optical and infrared wavelengths, imaging Fabry-Perot devices 
are used in three different ways: 
(i) to obtain a spectrum at each pixel position over a wide field 
by tuning the etalon,
(ii) to obtain a single spectrum of a diffuse source which fills 
a large fraction of the aperture (from one or more deep frames at the same
etalon spacing), and 
(iii) to obtain a sequence of monochromatic images
within a field defined by the Jacquinot spot 
(tunable narrowband imaging). Figure~\ref{differentModes}
summarises the different modes and the types of data they produce.
In this paper, we will concentrate on analysis techniques
adapted for final case, tunable filter imaging. However, many of the
methods presented herein are common to all three modes of operation.

Observations of a particular field consist of one of more
sets of images called a scan (or a {\sl stack} or {\sl spectral cube}).
A {\sl scan} is a series of images (or {\sl slices}) in which each one
has been taken at a slightly different plate spacing. This is achieved
by tuning the filter to a new passband between each exposure.
Ideally, scans should be repeated three or more times to permit
median-filtering of cosmic-ray events and ghost images from
slices of common wavelength. Removal of ghosts requires the telescope
to be offset between scans and the Fabry-Perot to be tilted to
ensure separation of ghosts and parent sources. 
In this situation, the ghosts will shift in the opposite direction
to the parent objects and by twice the amount.

At a given (constant) order of interference, a one-to-one
relationship exists between plate spacing and image wavelength.
At the telescope, the spacing is encoded in software units known as
{\sl Z-step values}, due to the movement of the Fabry-Perot
plates during both tuning and scanning.
Wavelength calibration frames take the form of long ``sausage-cubes'':
small images (few pixels square) at many wavelength settings
(typically 100 or more). The small image area permits expediency while
scanning the plates at sufficiently high sampling resolution.

\begin{figure}
\psfig{file=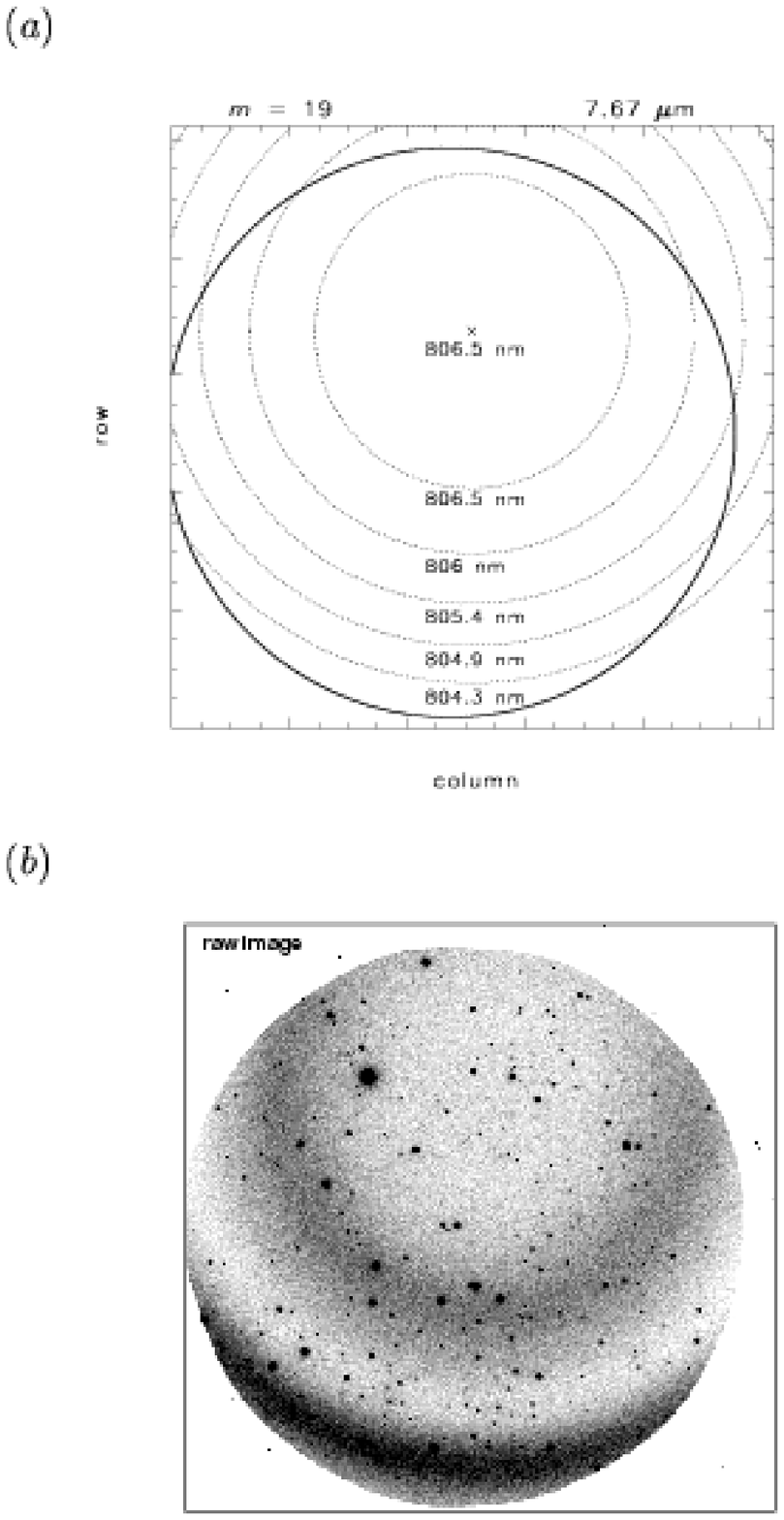,width=\hsize}
\caption[Contours showing field phase effect and associated night-sky rings]
{($a$) Contour plot showing the change in wavelength across
the field. In this example, the values at the top of the panel are
order of interference ({\em left}) and effective plate spacing 
({\em right}).
($b$) Raw image showing the ring pattern of
atmospheric night-sky emission lines resulting from a wavelength gradient
across the field. The wavelength contours in ($a$) match those of the image
in ($b$).  
}
\label{f:rawImg}
\end{figure}

If the Fabry-Perot is used in the pupil (as is usually the case), a 
{\sl  phase effect} shows itself as a position-dependent change in wavelength
across the field. Figure~\ref{f:rawImg}($a$) illustrates this with
data from the {\sc Taurus} Tunable Filter (TTF) system
(Bland-Hawthorn \& Jones~1998a,b) at the Anglo-Australian
Telescope (AAT). The centre of the
interference pattern is coincident with the {\sl optical axis} and
the distance of an object from this point is its {\sl optical radius}.
The phase change over the field gives rise to
{\sl night-sky rings}, broad diffuse circles of atmospheric 
OH emission lines on some images, centred on the optical axis 
(Fig.~\ref{f:rawImg}$b$). 
They are circular because the phase change makes for a specific
off-axis angle where the interference equation is
satisfied for a specific wavelength. For cases in which this effect 
is present, the flux needs to be re-normalised in radial bins, 
since it is not corrected through division by a flatfield.
If a tunable filter is used in the field ({\em i.e.}~after the field lens)
then the abovementioned effect does not occur.
Meaburn (1976) shows examples of this arrangement for Fabry-Perot
devices used in the near infrared.

The underlying principle of a tunable filter is the enlargement of
the Jacquinot spot at narrow plate spacings, and therefore, low spectral
resolutions.
The {\sl Jacquinot spot} is defined as the field about the optical 
axis within which the peak wavelength variation
with field angle does not exceed $\sqrt{2}$ of the etalon band-pass 
(Jacquinot~1954; Taylor \& Atherton~1980).
This angular field can be used to perform close to monochromatic imaging.


\section{Definitions}

\subsection{The Instrumental Response Function\label{response}}

The most direct route to the Airy function, the instrumental response 
of the Fabry-Perot,
is to use complex exponential notation. An incoming plane wave with 
wavelength $\lambda$ 
at an angle $\theta$ to the optical axis enters the etalon cavity and 
performs a series of internal reflections. 
If the highly reflective inner surfaces have reflectivities of 
${\cal R}_1$ and ${\cal R}_2$, we can
sum over the complex amplitudes of the outgoing plane waves such that
\begin{equation}
{\cal I} = 1 + {\cal R}_1{\cal R}_2 e^{2 i \delta} + {\cal R}_1^2 
{\cal R}_2^2 e^{4 i \delta} + .... = {{1}\over{1-{\cal R}_1 ,
{\cal R}_2 e^{2 i \delta}}}
\label{complexSum}
\end{equation}
in which $2\delta$ is the phase difference between successive rays. 
The total transmitted intensity is
proportional to the squared modulus of the complex amplitude or
\begin{equation}
{\cal A} = {\cal I}{\cal I}^* = {{1}\over{1+{{4{\cal N}^2}\over{{\pi}^2}} 
sin^2 2\pi\mu  l \lambda^{-1}\cos\theta}} ,
\label{airy}
\end{equation}
where the refractive index and the plate separation of the cavity are 
$\mu$ and $ l $ respectively, and 
${\cal N} = \pi\sqrt[4]{{\cal R}_1{\cal R}_2}(1-
\sqrt{{\cal R}_1{\cal R}_2})^{-1}$.

Clearly, the Airy function has a 
series of periodic maxima whenever
\begin{equation}
m\lambda = 2\mu l \cos\theta
\label{famous}
\end{equation}
which is the well-known equation of constructive interference in the $m$th 
order, with $\theta$ being the off-axis angle of the ray, (as subtended at
the etalon), and $\mu = 1$, since the Fabry-Perot is air-spaced. 
The quantity ${\cal N}$ is called the 
{\sl reflective finesse} and depends 
only on the values of ${\cal R}_1$ and ${\cal R}_2$.
It is normal procedure to manufacture an etalon with two identical 
coatings such that ${\cal R}_1 = {\cal R}_2$, but in App.~B we describe 
the extraneous etalon effect, where this condition is not met.

\begin{figure}
\psfig{file=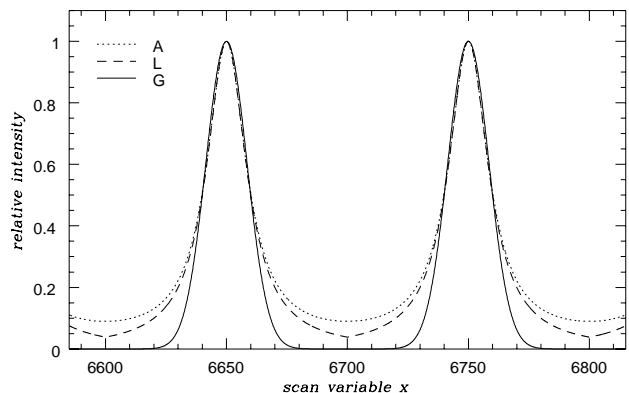,width=\hsize}
\caption[Three cyclic functions listed in Table 1 (column 2)]
{Three cyclic functions listed in Table 1 (column 2) shown at 
low finesse to emphasize their differences.}
\label{f:cyclic}
\end{figure}

\begin{figure}
\psfig{file=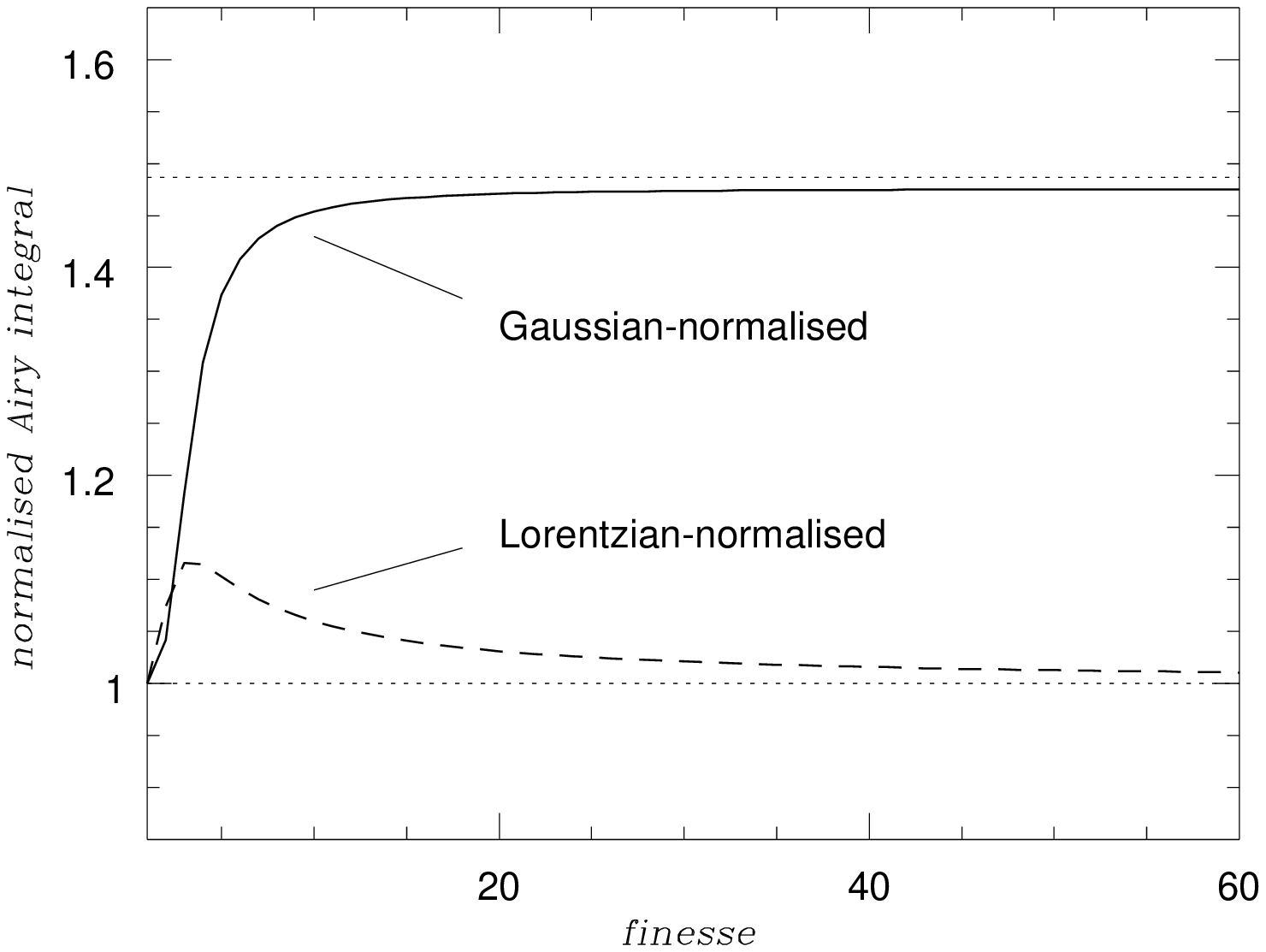,width=\hsize}
\caption[The integral of the Airy curve as a function of finesse]
{The integral of the Airy curve as a function of finesse 
normalized with respect to the Gaussian 
and Lorentzian integrals. The asymptotic limit of the upper curve 
is $\sqrt{\pi\ln 2}$.}
\label{f:finesse}
\end{figure}

In order to arrive at the correct calibration procedure, it is important
to understand the nature of the Airy function. 
Figure~\ref{f:cyclic} illustrates the three cyclic 
functions in Table 1 and shows that 
the area of the Airy function 
always exceeds the integral of the other functions for a given 
spectroscopic resolution. By analogy with 
the Lorentzian profile, the coefficient of the $\sin^2$ factor in 
Eqn.~(\ref{airy}) determines the width of 
the function. The quantity $\Delta x/\delta x$ is called the {\sl effective 
finesse}, where $\delta x$ is the spectral purity 
(defined in Eqn.~(\ref{purity}) below)
and $\Delta x$ is the periodic free spectral range. 
In Fig.~\ref{f:finesse} we illustrate how the Airy function, when 
normalized to the Gaussian and Lorentzian functions, depends on ${\cal N}$.
In practice, there are factors other than coating reflectivities which 
contribute to the effective finesse $-$
aperture effects, imperfections within the plate coatings, etc. $-$ 
some of which can serve to make the 
instrumental response more Gaussian than Lorentzian in form 
(Atherton~\etal~1981). 

Beyond a finesse of roughly 40, the Airy function is highly Lorentzian. 
The reason for this is clear when looking at 
how the Airy profile has been written in Table~\ref{t:cyclicfunctions}. 
At high finesse 
($\Delta x >> \delta x$), if we expand $x$ about the 
peak of the profile, the small angle formula reduces the Airy function 
to the Lorentzian form. The normalized Airy 
integral depends only weakly on finesse at large finesse values.

\subsection{The Jacquinot advantage\label{jacquinot}}

There are several approaches to deriving the Jacquinot advantage 
(Roesler~1974; Thorne~1988), {\em i.e.}~the throughput advantage
of the Fabry-Perot interferometer. 
By considering the solid angle subtended by the innermost ring and using 
the small angle formula with Eqn.~(\ref{famous}), 
we arrive at the important relation
\begin{equation}
\theta^2 = {2\over R} = {\Omega\over\pi},
\end{equation}
which assumes that $\mu = 1$ and that $\Omega$ and $R$ are the solid angle 
and spectral resolving power of the ring, respectively.

This equation has been used to demonstrate that Fabry-Perot interferometers, 
at a given spectroscopic resolution, have a much
higher throughput than more conventional techniques (Jacquinot~1954;~1960). 
But the Jacquinot relation has another 
important consequence. The solid angle of a ring defined by its FWHM 
intensity points can also be written
\begin{equation}
\Omega = 2\pi\theta\ \delta\theta = \pi\lambda ({\cal N} l)^{-1} .
\end{equation}
For a fixed etalon spacing, the solid angle, and hence the spectroscopic 
resolution, of all rings is a constant. This allows us
to write down a simple equation valid for all rings for the 
signal-to-noise ratio in a monochromatic unresolved
line, namely,
\begin{equation}
{\rm SNR} = s \sqrt{ 
\frac{ (\epsilon\tau\eta)
\left({\delta\lambda^{\prime}}\over{\delta\lambda} \right)
\left({\Omega\over\omega}\right) }
{ (s+b+ \frac{f\ \delta\lambda} {\delta\lambda^{\prime}\epsilon\tau} 
\sigma^2_R) }
}
\label{biggie}
\end{equation}
where $s$ and $b$ are the source and background flux (cts pix$^{-1}$ 
s$^{-1}$), $\epsilon$ and $\tau$ are the efficiency and 
exposure times respectively, $\omega$ is the solid angle subtended 
by a pixel. The quantity $\delta\lambda^{\prime}$ is the 
wavelength dispersion (\AA\ pix$^{-1}$) and $f$ is the number of CCD 
exposures combined to form the deep spectrum. 
The factor $\eta$ is discussed below.
We normally choose to place the ring center at one corner of the 
field for two reasons. First, it is often necessary to tilt the 
etalon in order to throw ghost light out of the field. 
Second, the factor $(\Omega/\omega)^{0.5}$ in Eqn.~(\ref{biggie}) now 
ensures that the spectroscopic sensitivity is constant over most of 
the field.  There will be an almost linear drop-off in sensitivity
at large off-axis angles (far corner from the optical axis) where the 
rings become seriously incomplete.

An important characteristic of a spectrometer is its {\sl spectral purity},
\begin{equation}
\delta x = \frac{\int_0^{du} {\cal A} (u) \; du}
{\int_0^\pi {\cal A} (u) \; du}
= \frac{2}{\pi} \tan^{-1}\Bigl(\frac{2{\cal N}}{\pi}\tan\frac{\pi}{{\cal N}}
\Bigr),
\label{purity}
\end{equation}
namely, the smallest measurable wavelength difference at a given 
wavelength (Thorne~1988). This
is usually defined as the intensity FWHM of the instrumental profile. 
When considering the amount of light transmitted by a spectrometer,
we need to consider the total area under the instrumental function. 
This issue is rarely mentioned in the context of long-slit 
spectrometers partly because their response is highly 
Gaussian,\footnote{The theoretical sinc$^2$ response of slit-aperture devices
is rarely achieved in practice.} in which case the `effective 
photometric bandpass' (total area divided by
peak height) is very close to the FWHM of the instrumental profile 
(see Table~\ref{t:cyclicfunctions}).

\begin{figure}
\psfig{file=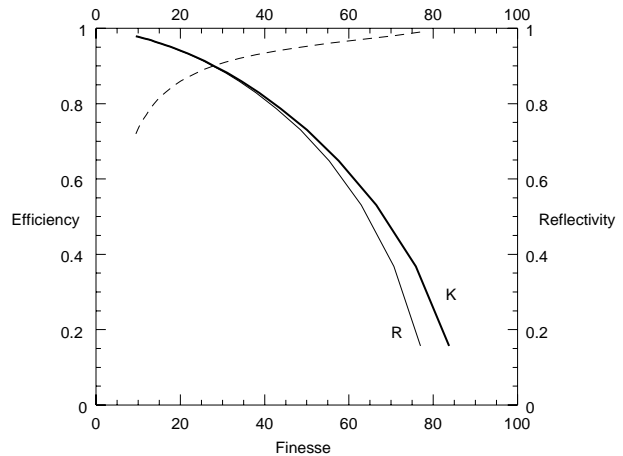,width=\hsize}
\caption[]
{A plot of efficiency against effective finesse for $R$ and $K$-band
Queensgate etalons. The $R$-band calculation is for a wavelength of 6700\AA\
and a plate flatness of $\lambda/150$ (after coating), measured at
6330\AA. The $K$-band etalon is for a wavelength of 2.2\um\ and
a plate flatness of $\lambda/50$. The curves decline less rapidly
if a higher plate flatness is achieved. We have included losses of 
0.2\% due to coating absorption and scattering; these tend to have
more effect at higher finesse. The {\em dashed} curve shows how the finesse 
depends on the coating reflectivity.
}
\label{finesse_eff}
\end{figure}

High spectral purity does not necessarily equate with high
efficiency. At values of low finesse, the efficiency is high 
but spectral purity is compromised by the amount of power in 
the wings, thereby reducing spectral definition (Fig.~\ref{finesse_eff}). 
Conversely, at high finesse a sharper
instrumental profile is marred by a loss in efficiency.
The optimal trade-off lies with values of finesse in the range
${\cal N} = 30$ to 40.

At such moderate finesse, the Airy function allows through 50\% more light 
than a Gaussian profile with
equal spectroscopic resolution (Fig.~\ref{f:finesse}). 
Thus, the {\sl effective photometric bandpass}, $\delta\lambda_{\rm e}$, 
is almost 60\% 
larger than the bandpass defined by the profile 
FWHM $\delta \lambda$. The factor $\eta$ in Eqn.~(\ref{biggie}) corrects for 
a calculation based on the FWHM of a ring 
and is defined as $\delta\lambda_{\rm e}/\delta\lambda$. Technically, 
$\delta\lambda_{\rm e}$ should be adopted as 
the spectral resolution of the Airy instrumental profile, otherwise 
we are forced to an inconsistency when comparing
Fabry-Perot spectrometers to other devices.

\begin{figure}
\psfig{file=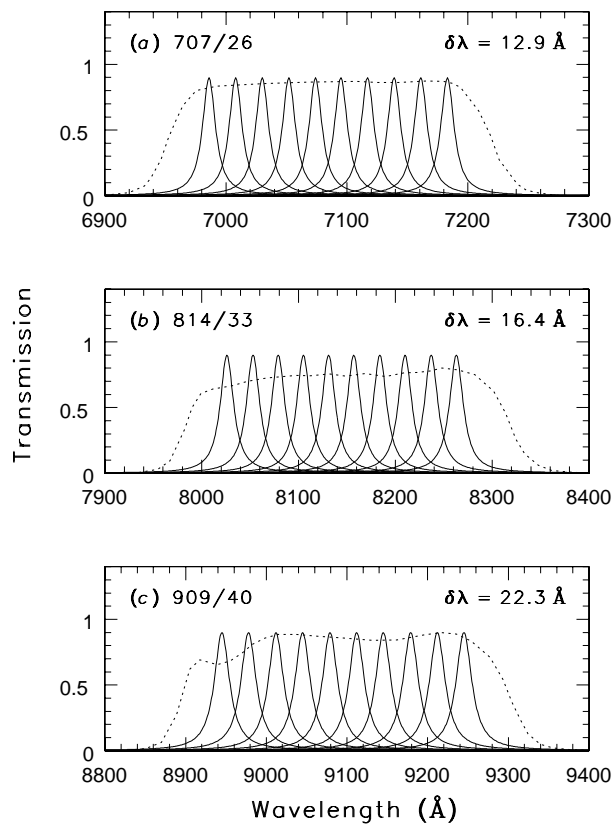,width=\hsize}
\caption[]
{Passband sampling used in the {\sl TTF Field Galaxy Survey} 
scans through ($a$) 707/26, ($b$) 814/33 and ($c$) 909/40~nm
blocking filters. The extent
of each blocking filter is also shown ({\em dotted lines}).
Note that the wavelength scale differs between panels.}
\label{figarrange}
\end{figure}

Using the Fabry-Perot as a tunable filter necessitates tuning the
bandpass to several discrete wavelengths and obtaining an image at 
each. Figure~\ref{figarrange} shows such an arrangement for the
{\sl TTF Field Galaxy Survey}, made with TTF at the AAT (Jones~1999).
Also shown in this figure are the filters used to block light
transmitted at unwanted neighbouring orders.  
The ideal spacing of bandpasses for such detection/photometry applications
is $\pi / 4 \times {\rm FWHM}$, which is half of the effective photometric
bandpass. However, for kinematic work, this sampling should be increased
to ${\rm FWHM}/2$.

\section{Data Reduction}
\subsection{Flatfield Correction}

It is hard to overstate the importance of the {\sl whitelight cube}. 
This is obtained by observing a whitelight
source over the same range of etalon spacings used in the actual 
observations. The whitelight cube
maps the response of the filter as a function of position and 
etalon spacing. There are three effects that
we wish to divide out from the data. Firstly, the narrowband filter 
response, when convolved with the 
instrumental response, leads to a modulation in the observed 
spectrum. Secondly, it is well known that
filters have variable responses in both collimated and converging 
beams (Lissberger \& Wilcock~1959). Thirdly,
we seek to remove any inhomogeneities in the filter structure 
as a function of position. When the
whitelight cube is compressed in the spectral dimension, it 
provides a very high signal-to-noise flatfield
for removing pixel-to-pixel sensitivity variations.

Several approaches exist for the creation and use of whitelight 
flatfields, depending on the exact nature of the data. 
Dome flats do not mimic the sky well in cases where
light scatters off the dome and irradiates the detector at 
extreme angles. This can fill in the whitelight response structure
with wavelength. Alternatively, twilight
flatfields contain Fraunhofer lines which generate fringe structure 
not present in science data. The underlying point is that 
both the {\sl spatial} and {\sl spectral} distribution 
of the flatfield source
must be well-understood for its proper use in removing 
pixel variations.

Final flatfields can be constructed from raw flatfield frames in one 
of two ways, dependent upon whether interference fringes are present in the
CCD used. 
Such fringes are due to interference within the substrate of the
CCD at red wavelengths and are unrelated to the interference 
within the Fabry-Perot. In the absence of fringing the
flatfield frames can be normalised and retained as individual components
of the scan, and used to divide images on an individual basis in the
usual way. If fringing is present in both the flatfields
and science frames, division of the latter by the former can be tried,
but its success relies on the stabilty of the fringe pattern. Typically,
Fabry-Perot plate spacings are not sufficiently stable to produce fringe 
structures constant over the hours that pass between
flatfields and science frames. Alternatively, a fringe
correction frame can be created by combining the science frames at a 
given etalon setting, using median filtering to remove the objects in the 
field, (assuming that the object position has been dithered between 
exposures). This produces a final frame of the fringe pattern alone, 
which can be subtracted. However, the same technique can not be used for
removing fringes from the flatfield. More elaborate operations can be
adopted, usually at the expense of added noise to the frame cleaned of
fringing. In the end one has to determine whether it is better to use
a noisier frame cleaned of fringes, or a less noisy frame on which the
fringes remain. This depends on the specific science requirements of the 
data.

While flatfields (used in either manner) successfully remove
pixel-to-pixel variations, the instrumental response across the beam
is modulated by the phase effect. In particular, care must be exercised
in the use of flatfields near the edges of the blocking-filter
transmission profile, as the effectiveness of the flatfield then
becomes a function of optical radius. 
In the case of photometry of point-source objects, we can account 
for any residual radial changes in sensitivity 
by photometrically registering objects
through a fit with optical radius 
at a later stage (Sect.~\ref{s:ELcandExtract}).

\subsection{Night-Sky Ring Removal \label{holes}}

For detection of faint point-sources, 
night-sky rings need to be removed so that the detection 
software can successfully fit the sky background,
(which is typically done through the use of a two-dimensional sky mesh.) 
Although the rings look circular, they are in fact slightly elliptical 
(0.5\%) due to the imperfect alignment of the CCD plane and
optical distortions within the focal reducer.

Glazebrook \& Bland-Hawthorn (2001) have
demonstrated that the night sky lines can be removed from
data to the level of 0.03\% or better using nod and shuffle
techniques. Here, the telescope is nodded between the object
and sky positions in synchrony with charge being shuffled on
the CCD with a cycle time of, say, 60 secs. This results in a 
50\% loss of observing efficiency but does in fact lead to superb
sky subtraction. Nod and shuffle with TTF produces almost perfect
residuals largely because there is no entrance slit, and the
monochromatic band is sampled by hundreds of CCD pixels along
the optical radius vector.

In those cases where nod and shuffle is not available or not feasible, 
alternative ways of removing the sky can be followed. 
If the sky background and night sky lines can be taken as having a fixed
pattern across the field of view -- usually a safe assumption for the
relatively small field and long exposure times of the TTF -- the
problem of sky subtraction can be most readily approached by
exploiting the azimuthal symmetry of the sky component relative to the
optical axis. A flat sky will be evident in the data as a radial
variation in the background, with night sky lines appearing as broad
rings. To parameterize the spectrum of the sky, the basic method is to
azimuthally bin the image about the optical axis.  Bright stars and
regions of evident emission are masked from the image prior to the
binning procedure. The sky spectrum is then calculated from the masked
image using a median combination of the pixels remaining at each
radius. The median will eliminate fainter emission that the mask may
have missed and additional cosmic rays.

The position of the optical axis is crucial to determining
the accurate phase error over the field. There are several
ways to find its location:
\begin{enumerate}
\item If the tunable filter can be tuned to an interference order
$m > 70$ or so, the monochromatic ring will be narrow
enough  to be fit with least-squares techniques. Specific
procedures for fitting the rings are discussed below and in App.~A.

\item We can generate a well-defined ring by differencing two
low order monochromatic scans where one is tuned slightly in
$Z$-space ({\em e.g.}~half a bandpass) compared to the other. 
The residual produces a P-Cygni profile along the optical radius
passing through zero counts which can then be fit with standard 
procedures discussed below and in App.~A. This residual image is typically 
very flat and not susceptible to CCD bias, flatfield or response 
variations.

\item Diametric ghosts can be very useful for identifying
the optical axis since these, by definition, reflect about the
optical axis. One technique is to generate the ghost map by
imaging through a mask of pinholes arranged in a regular grid.
At the AAT we use a matrix mask consisting of 100\um\ holes 
drilled every 1~cm. 
The optical axis can then be identified as the crossing point for 
all vectors which join the pinhole images with their respective 
ghosts (Fig.~\ref{matrix}).

\begin{figure}
\psfig{file=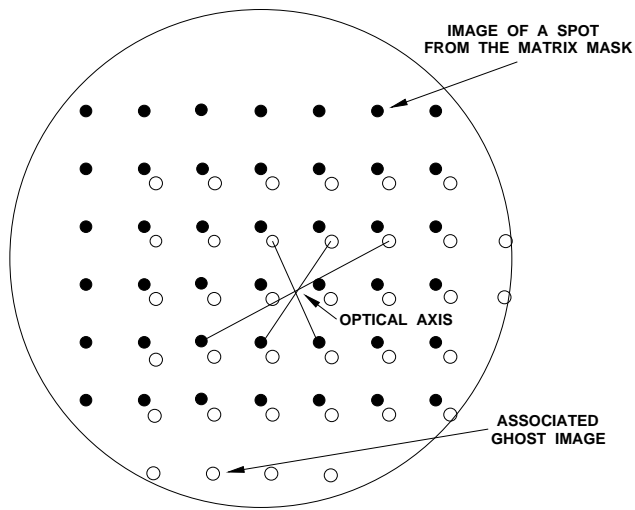,width=\hsize,angle=270}
\caption[Use of a pinhole mask image to find the optical axis]
{Use of a pinhole mask image to find the optical axis.
The ghost image of each pinhole is diametrically reflected about the
optical centre. The intersection of two or more lines joining
pinhole images to their corresponding ghost images will give
the location of the optical axis.
}
\label{matrix}
\end{figure}

\item A more involved approach is to derive the phase surface
by scanning the TTF over a strong emission line (Bland \& Tully 1989). 
Each spectrum within the monochromatic cube must
now be fit with Lorentzian profiles. The resulting surface is 
highly paraboloidal and the optical axis along its centre is
well defined. 
\end{enumerate}

\begin{figure*}
\psfig{file=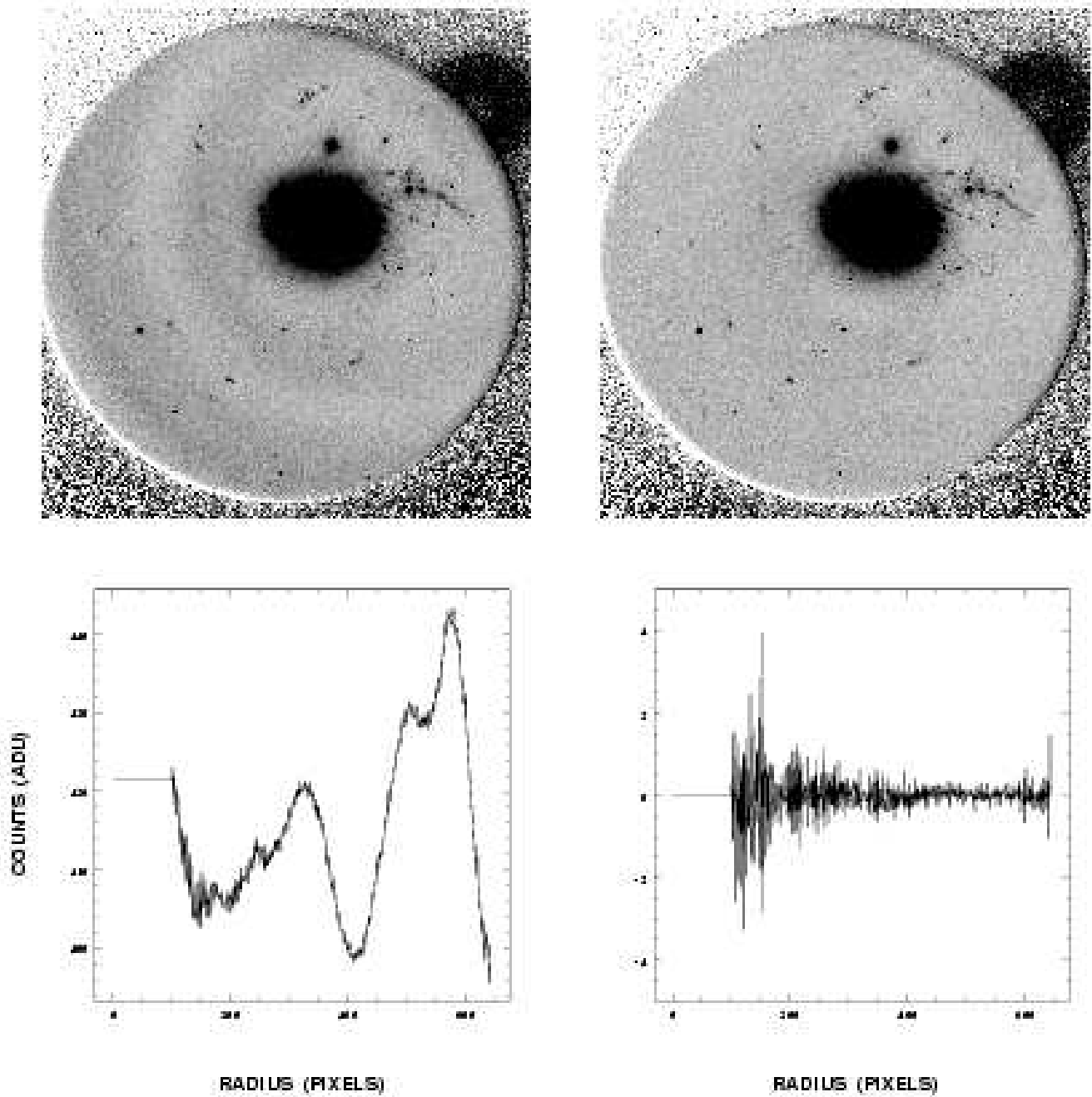,width=\hsize}
\caption[Azimuthal sky removal]
{Example of night-sky ring removal using azimuthal fitting and
subtraction. ({\em Left}) Sky ring pattern on the original image
with a cross-section ({\em below}) showing the amplitude of the variation.
({\em Right}) The same image and cross-section following subtraction.
Note that the remaining faint background structure in this case is real.
}
\label{patfig}
\end{figure*}

A quick method for finding the approximate centre of the night-sky 
ring pattern is to solve the equation of a circle from three points.
However, for the purpose of subtraction, circular fitting 
is usually insufficient and more sophisticated strategies must 
be adopted (App.~\ref{ringFitting}).
Once the sky spectrum has been
determined, a two-dimensional sky ``image'' can be constructed by
sweeping the sky spectrum azimuthally around the position of the
optical axis, in effect reversing the binning procedure. This sky
image may then be subtracted from the data image. As is illustrated in
Fig.~\ref{patfig}, this procedure works quite well for sparse 
fields, where only a small percentage of the pixels need to be masked 
from the binning calculation.
We note that this method works best if one gives some consideration to
the placement of the optical axis at the time of observation:
within the field of view but away from the object of interest.
Placement should also ensure that bright ghosts in the field are avoided.

In those cases where the field is densely populated,
the empirical removal of a uniform azimuthal sky component may not
be feasible. At this point, the sky rings must be individually fit
and parametrized.
The core of the fitting problem is that the sky-rings are typically
incomplete, passing beyond the field-of-view for part of their length.
Defining the problem as minimising the sum of squares, the
so-called geometric distance, is computationally very expensive
and requires careful weighting schemes and application
({\em e.g.} Joseph 1994).  After years of experimentation with conventional 
methods ({\em e.g.} Bland-Hawthorn, Shopbell \& Cecil 1992), 
we have settled on algebraic distance methods as the most 
economical and reliable approach to fitting incomplete arcs, 
as described in App.~\ref{ringFitting}. 
One persistent problem has been the geometric distortion of Cassegrain 
focal reducers exacerbated by slight tilts in the alignment of the 
CCD plane.

An alternative method for removal is feasible in the case where 
the objects of interest are much smaller than the
ring structure ({\em e.g.}~at very low spectral resolution when the
night-sky rings are large and diffuse).
The rings are removed by treating each science exposure individually.
A background map is created by median-filtering copies of the original,
each one offset in a regular grid pattern from the other by a few pixels.
The result is then smoothed and subtracted from the original, 
leaving little or no night-sky residual.
The method depends upon the rings being of lower spatial frequency than the
objects, and as such, would be less successful with extended sources. 
Bland-Hawthorn \& Jones~(1998b) demonstrate this
technique with an example.

\subsection{Wavelength Calibration}

The tunable filter is wavelength-calibrated through {\sl scans} consisting
of many images taken at a series of gap spacings. Spacing through the
scan is controlled in software through a parameter, $Z_s$, of arbitrary
units and proportional to physical plate spacing.
Ordinarily, two scans are required: 
an {\sl emission-line scan} of an arc lamp such as copper-argon
or neon, and a {\sl continuum scan} of a white-light source 
such as a tungsten lamp (the white-light calibration cube), as
mentioned earlier. 

It is essential that both types of source are diffused by suitable
optics. Past experience has shown that anomalous phase effects are 
produced by incorrect use of calibration lamps. They must 
illuminate the instrument evenly and should not be placed between the 
telescope focus and the CCD.

An additional consideration in the choice of a suitable arc lamp
is that a minimum of two lines, (preferably more), should be
visible within the restricted wavelength coverage of the blocking filter
(typically 25 to 30~nm). Lamps in use with TTF system at the 
AAT are shown in Table~\ref{t:lamps}. Around 500~nm, the
astrophysical lines of \Hb\ and \Oiii\ found in planetary nebula
are a suitable alternative, and Acker~\etal~(1992) list many suitable
objects.

\begin{figure}
\psfig{file=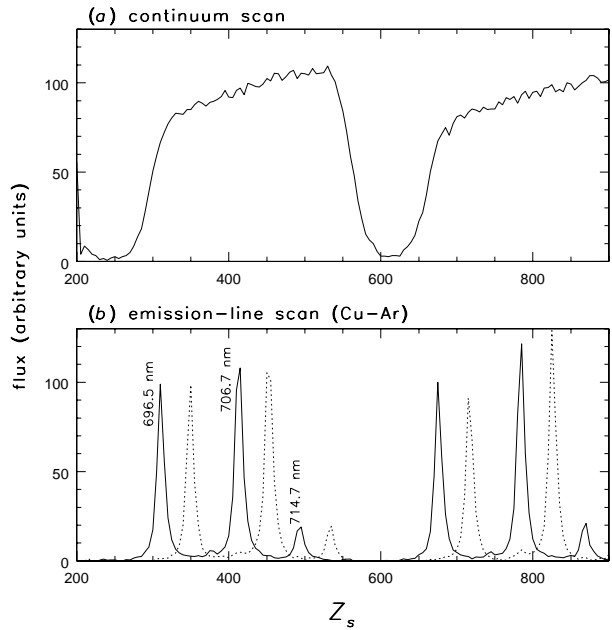,width=\hsize}
\caption[TTF calibration scans]
{Example tunable filter calibration scans through the same blocking
filter, showing ($a$) continuum, and ($b$)
emission-line spectra. The two scans in ($b$) were measured
on-axis ({\em solid line}) and off-axis ({\em dotted}).}
\label{f:lampScans}
\end{figure}

The continuum scan (Fig.~\ref{f:lampScans}$a$) is used to check where
(in $Z_s$) the orders of interference occur and to ensure no overlap.
Continuum scans can be sampled quite coarsely 
($\Delta Z_s > 20$) and should cover a range of more than one order,
since the location of these is not known at the outset.
Overlap of the resulting blocking filter traces indicates a plate spacing
that is too large, 
causing the sampling of more than one order of interference.
Once two orders have been located, 
a well-sampled scan ($\Delta Z_s \sim 3$) through
an arc lamp can follow (Fig.~\ref{f:lampScans}$b$). 
Such a scan covers more
than one order so that at least one arc line appears in adjacent 
orders. This is necessary for a determination of physical plate spacing.
Typically the number of arc lines available over the wavelength coverage
of the blocking filter is few. The two profiles in Fig.~\ref{f:lampScans}($b$)
show the wavelength offset between on and off-axis positions.

Lines observed in the emission-line scan are identified both by the 
wavelength and $Z_s$ at which they occur. Across a single 
order, the $(Z_s,\lambda)$ relationship is linear,
\begin{equation}
\lambda (Z_s) = C_1 Z_s + C_2 .
\label{e:lambdaVSz}
\end{equation}
To determine the plate-spacing in physical units (such as microns) we need to
first determine the order of interference. 
If $\Delta Z_0$ is the free spectral range (in $Z_s$-units) separating 
an arc line (wavelength $\lambda_0$) in adjacent orders,
then the order of interference $m$ is given by
\begin{equation}
m = \frac{\lambda_0}{\Delta Z_0 \cdot C_1} .
\label{e:order}
\end{equation}
With $m$ known, the spectral resolving power $R$ and passband width
$\delta \lambda$ can be determined through
\begin{equation}
N_{\rm R} \cdot m = \frac{\lambda_m}{\delta \lambda} = R.   
\label{e:resolvingPower}
\end{equation}
The large scan range of TTF sees resolving powers anywhere in the range
of 150 to 1000.
The physical plate-spacing calibration follows as an adaptation
of Eqn.~(\ref{e:lambdaVSz}), namely
\begin{equation}
L = \frac{m}{2 \cos \theta_{\rm off}} \cdot \Bigl (C_1 Z_s + C_2 \Bigr ),
\label{e:spacing}
\end{equation}
where $L$ is the effective plate spacing and $\mu = 1.00$ for air.
By {\sl effective} plate-spacing, we mean the gap as measured parallel to the
optical axis, irrespective of etalon tilt. The {\sl true} gap between the
plates is smaller than this, being $L \cos \alpha$, where
$\alpha$ is the tilt of the etalon,
(typically a few degrees to deflect ghosts out of the beam). 

At very narrow gaps ($\simlt 3$\um),
there is a wavelength-dependent phase-change in the reflections between
optical coatings on the inner plate surfaces. This introduces a
degeneracy that requires an additional term to be introduced into
Eqn.~(\ref{famous}). The effect can be measured at such narrow plate-spacings, 
provided the phase-change behaviour of the inner coatings has been
determined through independent means (Jones \& Bland-Hawthorn 1998).

The angle $\theta_{\rm off}$ in Eqn.~(\ref{e:spacing}) 
is that separating the rays on-axis and those sampled for calibration.
The wavelength offset between both points must be included in the
calibration, through relations derived in
Bland-Hawthorn \& Jones~(1998a,b). For example,
suppose that the point used on the detector is offset by 
500~pixels from the optical axis. If the focal length of the camera
is 130~mm and detector is a $1024 \times 1024$~pixel CCD of 
$25 \times 25$~mm, then 
\begin{equation}
\theta_{\rm off} = \frac{500}{1024} \times \frac{25}{130} = 0.094\,\,\,{\rm radian}.
\end{equation}
If the observing conditions cause measurable drifts in wavelength
of the etalon bandpass one can treat the affected frames as 
wavelength-averaged over the exposure time and bandpass from
the time dependence. Temporal variations can be measured through
rapid scans (either small pixel-area or charge-shuffle) and fit by
a smoothly-varying function.

\subsection{Image Alignment}

\begin{figure}
\psfig{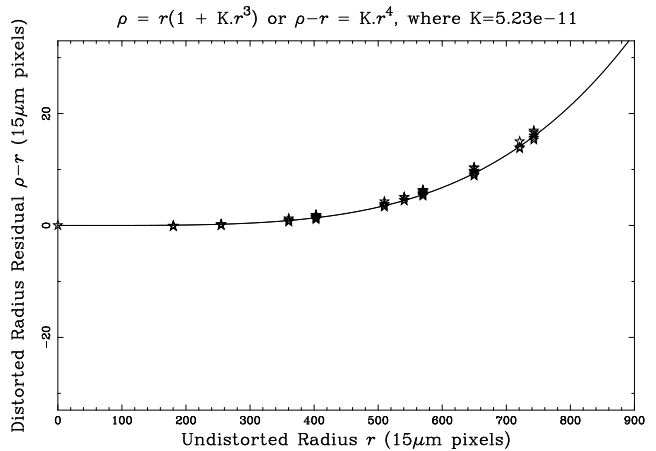}
\caption[Radial distortion in the \taurus2\ field]
{Radial distortion in the \taurus2 focal reducer of the AAT, 
as measured across the focal
plane by C.~Tinney using an MIT-LL CCD (15\um\ pixels). 
The residuals between the distorted ($\rho$)
and undistorted ($r$) radii have been plotted and fit with the power
law shown at top.}
\label{radial}
\end{figure}

The offsets introduced for ghost-image removal necessitate spatial registering
of scans. 
There is distortion present in the field extremities of many focal
reducers used to accommodate Fabry-Perot instruments, including 
the \taurus2\ focal reducer on the AAT.
To measure this, we use a matrix mask 
consisting of 100\um\ holes drilled every 1~cm. 
Figure~\ref{radial} shows a measurement of this by C.~G.~Tinney 
through the residual of the distorted radial distance 
($\rho$, as measured at the detector) against that
undistorted ($r$). At the largest radii the maximum distortion is 2\%,
and relative astrometry to better than 0.1 pix (0.03\as\ at f/8) is 
possible. The radial correction function requires terms up to $\rho^4$.
Scan images need not be corrected for this provided
the dithering is small ($\simlt 10$\as). 
Bacon, Refregier \& Ellis~(2000) give an elegant treatment of
the general solution to image distortions across a field
(their Sect.~5), which can also be applied to tunable filter data
in which high spatial precision is required.

\subsection{Variations in Seeing}

Typically it takes several hours to complete scans and
as such, the observations are susceptible to changes in seeing
between individual frames. In such cases it is necessary to
create a second version of the scan, in which all images have been 
degraded to the mean seeing $\langle W_{\rm w} \rangle$ of the worst 
frame in the original scan. This is done by covolving each original
frame with a gaussian kernal of FWHM
\begin{equation}
\langle W_{\rm k} \rangle = \sqrt{{\langle W_{\rm w} \rangle}^2 -
{\langle W_0 \rangle}^2},
\label{e:kernal}
\end{equation}
where $\langle W_0 \rangle$ is the mean seeing FWHM of the given frame. 
Both the convolved and unconvolved (original)
versions of the scans are used for the subsequent analysis.

\subsection{Co-addition of Image Stacks\label{s:coaddition}}

For the best detection and photometry of a field, 
there should be three or more
scans (as separate measurements), 
with slices $S(k,\lambda_1)$, $S(k,\lambda_2)$, \ldots, 
$S(k,\lambda_n)$, where $n$ is the number of slices per
scan and $k$ is the scan number (up to the maximum $K$ scans taken).
Each image corresponds to its central wavelength $\lambda_i$, 
($1 \leq i \leq n$).
There should also be matching frames $G(k,\lambda_1)$, \ldots, 
$G(k,\lambda_n)$,
that have been smoothed to match the worst seeing image.
A final co-added narrowband scan can be obtained by combining each
set of narrowband frames of common wavelength.
With two versions
of all frames (unconvolved and convolved), and two methods of combination
(straight summation or median-filtering), three versions of the
co-added scan are derived:
\begin{enumerate}
\item Straight summation of the {\sl convolved} frames at each wavelength,
\begin{equation}
\sum_{k=1}^{K} G(k,\lambda_i), ~~~~~~~~~\forall \; i=1, \, \ldots, \, n,
\end{equation}
\item Straight summation of the {\sl unconvolved} frames at each wavelength,
\begin{equation}
\sum_{k=1}^{K} S(k,\lambda_i), ~~~~~~~~~\forall \; i=1, \, \ldots, \, n,
\end{equation}
\item Median-filtering of the {\sl unconvolved} frames at each wavelength.
\end{enumerate}
Median-filtering cleans the cosmic-rays and ghost images from scan (iii). 
However, it also lowers the signal-to-noise ratio by 
30\% or more on the faintest objects, thereby making scans (i) and (ii)
more sensitive. Scans (ii) and (iii) have the better spatial resolution
but are inferior to scan (i) for seeing stability, and therefore, 
photometric integrity. Optimally, scans (ii) and (i)
can be used for the object detection and photometry, 
while (ii) and (iii) are used to build cosmic-ray
and ghost image catalogues.


\section{OBJECT DETECTION AND CLASSIFICATION}

\subsection{Initial Object Detection and Photometry\label{apertselect}}

Object detection and photometry on a stack of narrowband frames 
can be performed using any of the common software packages
for detection and photometry of objects, such as
\sext\ (Bertin \& Arnouts~1996) or FOCAS (Valdes 1993).
The prescriptions that follow were developed using \sext.
However, FOCAS has also been used with similar tunable
filter data ({\em e.g.}~Baker~\etal~2001).

The \sext\ algorithm initially sees 
images background subtracted 
and filtered with a convolution mask to optimise detection
sensitivity. Candidate objects are found as connected pixels above
the prescribed threshold and passed through a deblending filter
to separate into nearby overlapping objects if this is the case.
\sext\ accepts input in the form of one (or two) FITS images and several
files. We can make use of the dual-image capability to detect objects on the
straight-summed/unconvolved frames and then photometer these positions
on matching straight-summed/convolved frames. Basic shape and
positional properties are measured from the detection image while
photometric parameters are obtained from the second.

In the analysis of a full scan, the detection threshold should 
be set sufficiently low that all objects are recovered
throughout the stack, including some noise peaks. 
This ensures the faintest detections are not lost
due to variations in background through the scan. 
The object-finding software can be run
on each scan frame in turn, the catalogues of which 
can be concatenated into a full {\sl raw catalogue}. Each entry of this
is a single detection (or {\sl observation}) in 
$(x,y,N)$-space, where $N$ is the image containing the detection at
spatial coordinate ($x,y$).

As with all imaging surveys, an inherent problem is not knowing 
the size of each object before its detection. It is possible to 
use a series of apertures of increasing size to cover the
range of expected object sizes.
If the innermost aperture
is set to 1.35 times the seeing, the signal-to-noise
ratio is maximised for a spatially unresolved source 
(gaussian profile) in a background-limited noise regime. 
For the purpose of maximising signal-to-noise ratio, it is preferable for a
non-optimal aperture to be too big rather than small.

\subsection{Extraction of Object Spectra\label{s:extraction}}

It is necessary to 
group multiple wavelength observations of the same object
together. A straightforward algorithm for processing the
raw catalogue is as follows. The first observation
is written to the {\sl object buffer}, an array to store all observables for
a given object from each slice. The remaining lines of the raw catalogue
are then searched for $(x,y)$-matches to the object in the buffer. 
If all observations for the object are found, or the limiting search
depth is reached, the contents of the object buffer are summarised
and written to file. 
The next observation
is written to the cleared object buffer and the process repeats.
Depending on the nature of the fields observed, the raw catalogue can 
contain $\sim 10^4$ observations for a 10 slice scan.
Each time the object buffer is complete, the contents should be checked 
for at least two observations before being considered as an ``object''.
Such a double-detection criterion provides a robust way of removing 
noise-peak and isolated cosmic-ray detections in an automated fashion.
However, it also selects against emission-line detections in a single-frame.
Given the difficulty inherent in distinguishing such detections from
cosmic-rays and noise peaks without manual inspection, 
single-frame detections are more easily extracted separately.

Each double-detection object can be assigned an ID number and 
the distance of the object from the optical axis calculated and 
used to correct calibrated wavelengths for phase shift 
(Bland-Hawthorn \& Jones~1998b). 
The signal-to-noise ratio can be determined for the full rage of aperture 
sizes and those from the optimal aperture retained. Below-threshold
detections can be discarded.
If the final catalogue contains mainly 
galaxies, then subset catalogues for bright stars should also be created. 
These are needed for photometric registering during emission-line 
candidate extraction (Sect.~\ref{s:ELcandExtract}).

\subsection{Cosmic Ray and Ghost Image Removal\label{s:cosmicRay}}

At this stage in the reduction, most of the cosmic-rays present 
on the images should have been caught 
by the detection software.
The minimum two-detection criterion also
filters out those that the software misses. 
However, two classes of spurious object are missed by
both methods. Firstly, a cosmic-ray that occurs close enough to a
real object will avoid removal if not detected by the software as a 
separate entity. Left uncaught, it will appear as a flux change in
the nearby object, with very similar characteristics as genuine
emission-line flux on continuum. Secondly, ghost images 
are usually admitted by detection algorithms as real objects
because they show little change in flux or position between the frames of
a single raw scan. 
It is only through a frame-by-frame comparison with
the {\sl cleaned} images that they become apparent. 
Appendix~\ref{ghostfamilies} discusses the different types of ghost 
images and strategies for minimising their occurrence at
the outset, before they become part of the data.

Such spurious detections can be handled through the compilation of a
catalogue of all cosmic-ray and ghost-affected pixels in 
$(x,y,N)$-space, and cross-checked with the $(x,y,N)$ object list.
Any matches in the latter can then be removed.
The straight-summed and median-filtered scans 
(scans (ii) and (iii) of Sect.~\ref{s:coaddition}) can be searched 
by taking each matching pair of images and computing the ratio image
\begin{equation}
C(x,y) = \frac{S(x,y) + f_1}{f_2 \times M(x,y) + f_1}.
\label{e:cosRatio}
\end{equation}
Here, $S(x,y)$ and $M(x,y)$ are a pair of straight-summed and
median-filtered images, $f_1$ is an offset that governs search sensitivity
and $f_2$ is the ratio of the two images measured from stellar fluxes.
Spurious pixels are those for which $C(x,y)$ exceeds
$C_{\rm thresh}$, the {\sl cosmic-ray threshold}, which is sensitive to
$f_1$. The parameter $f_1$ should be fixed 
100 and 500 ADU) 
while $f_2$ and $C_{\rm thresh}$ 
vary freely with each image $N$. Trial divisions can be used to
optimise these parameters.
The final cosmic-ray/ghost catalogue is used in the emission-line
search described in Sect.~\ref{s:ELcandExtract}.

In the case where only a pair images exists (and median-filtering is not
possible), the approach of Cianci (2002, in prep) can be used to identify
affected pixels. The dithered images are registered and pixel fluxes compared as
a graph between the two images. In regions where they are identical, the fluxes
follow a $y = x$. Pixels that differ, whether affected by ghosts or cosmic rays,
deviate from this straight line and those beyond some threshold can be removed.
Alternatively, Rhoads~(2000) describes a convolution filtering-algorithm for
cosmic-rays, using the image point-spread function minus a delta function 
as its kernel. This method is also useful when there are too few
frames for median filtering.
Appendix~\ref{weakcosmics} describes how to test the effectiveness of
a given cosmic-ray removal technique. This is particularly valuable
in the case of data containing hundreds of faint events.

\subsection{Emission-Line Candidate Extraction \label{s:ELcandExtract}}

Telluric absorption features (Stevenson~1994), that give rise to 
variations in atmospheric transparency with wavelength, affect the
near-infrared spectral regions sometimes scanned by a tunable filter.
Therefore, before any emission-line candidate extraction takes place, 
it is necessary to register all object fluxes on to a common 
scale at an unaffected wavelength. The normalisation coefficients can be 
calculated using a sub-sample of stellar fluxes from the field.
A linear fit of flux ratio versus off-axis radius
accounts for off-axis changes not removed by the original flat-fielding.

The extraction of emission-line candidates then proceeds in three stages. 

\begin{enumerate}
\item An initial search is made for all objects detected in two
adjacent bands and no others. Such objects have no detected continuum on
the remaining frames.
A cross-check can be made between these and the cosmic-ray/ghost
catalogue (Sect.~\ref{s:cosmicRay}) and matches in $(x,y,N)$-space discarded.
The catalogue of non-continuum detections can then be set aside.
\vspace{5mm}

\item The second stage is the search for objects with an 
emission-line superposed on continuum.
Each object with a sufficient number of observations ({\em e.g.} five)
can have a straight line
iteratively fit to its continuum, initially with no rejection. 
The root-mean-square (RMS) scatter $\sigma$ about the line
can then be compared to the mean of the flux measurement
uncertainties, $\langle \Delta F \rangle$. The larger of the two
is taken as the
dominant source of uncertainty, $\sigma_{\rm dom}$, against which all
subsequent deviations should be measured. 
If $\sigma_{\rm dom} = \langle \Delta F \rangle$, successive fits 
should be weighted by the $1/\Delta F^2$ measurement errors; otherwise, fits
are best left unweighted. Points can then be rejected if they
deviate by more than $\sigma_{\rm dom}$ {\sl above} the continuum. 
A limit needs to be placed on the total number of points allowed
to be deviant ({\em e.g.} two), provided a sufficient number
of points remain to constrain the continuum fit. Identical reasoning
applies to the rejection of deviations {\sl below} the line,
(as in searches for absorption features), which we do not go into here.
Objects can be retained as emission-line objects if (1) the deviation
is greater than the chosen threshold above the line, and (2) a cross-check
with the cosmic-ray/ghost catalogue reveals no matches.
\vspace{5mm}

\item The final stage is processing of the raw object-detection  
catalogues in a search for object detections on a single frame. 
Cross-checking between these detections
and the cosmic-ray/ghost catalogue is made in the same way as for the
double-detection objects. However, the final catalogues can be too large 
(several hundred such objects per field) to inspect manually in the same
fashion. Many of the single-frame detections
are noise-peaks arising from the low detection threshold deliberately
employed for object detection.  
Therefore, the flux threshold cut applied to the 
single-frame detections should be determined from the double-detections. 
\end{enumerate}

\begin{figure*} 
\centerline{\epsfxsize=140mm\epsfbox{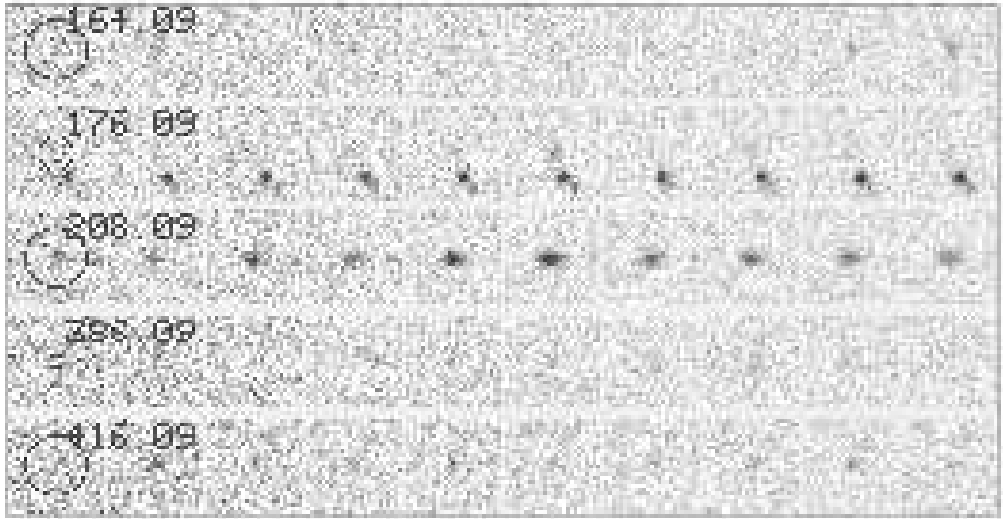}}
\vspace{0.5cm}
\centerline{\epsfxsize=80mm\epsfbox{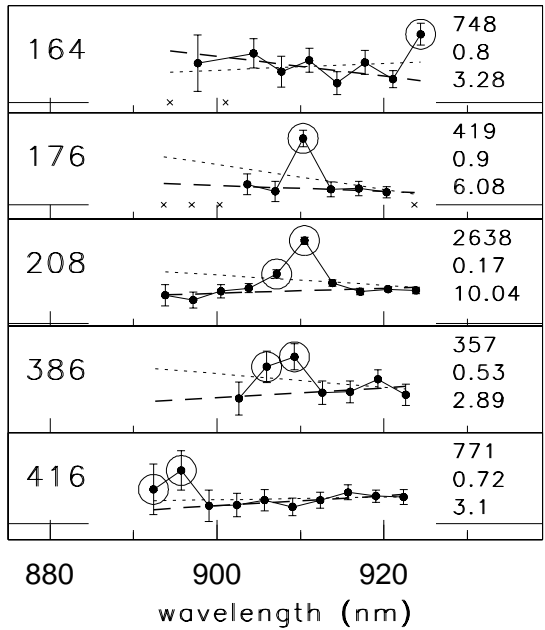}}
\caption{
{\em Top.} Strip-mosaic scans showing candidate examples from a 
deep galaxy survey field, scanned with the {\sc Taurus} Tunable Filter
(25\AA\ passband) in conjunction with a 909/40~nm blocking filter. 
Circles denote aperture size and the numbers are object labels.
{\em Bottom.} Narrowband fluxes for the same candidates. 
Numbers ({\em right}) are
flux (in ADU), star-galaxy parameter and emission deviation (in $\sigma$).
Also shown are the initial ({\em dotted}) and final ({\em dashed line})
linear fit to the continuum flux. Deviant points rejected from the
final fit are indicated ({\em circles}), as are non-detections in a
frame ({\em crosses}). The catalogue numbers ({\em left}) 
correspond to those in panel ($a$).}
\label{f:inspect}
\end{figure*}

Upon completion, all three catalogues of non-continuum and continuum-fit
objects can be combined.
Figure~\ref{f:inspect} shows examples of distant emission-line galaxies
detected for the {\sl TTF Field Galaxy Survey} (Jones \& 
Bland-Hawthorn 2001). Both the strip-mosaic (Fig.~\ref{f:inspect}$a$) 
and spectral scan (Fig.~\ref{f:inspect}$b$) of each object are
shown.


\section{FLUX CALIBRATION}

When we build up a data cube or take a series of observations, it is 
essential to think of the scan variable as 
the etalon gap $l$ rather than wavelength. 
The wavelength range is moderated by the filter;
the etalon gap is not. The physical plate scanning range is 
$l_0 \pm 2\Delta l$ $=$ $l_0 \pm 2(\lambda_0/2)$ where $l_0$ is the 
zeropoint gap and $\Delta l$ is the 
free spectral range in physical gap units. With this important distinction in
mind, for the flux in a standard star observation, we are able to write
\begin{equation}
S(l) = \int F_S(\lambda) {\cal A}(\lambda,l)\ d\lambda
\label{integral1}
\end{equation}
where  $F_S(\lambda)$ is the product of the stellar spectrum and the filter 
response. 
The limits of the integral in Eqn.~(\ref{integral1}) are defined 
by the bandpass of the entrance filter.\footnote{  
The transform is some form of a convolution
equation in that ${\cal A}(\lambda,l)$ broadens $F_S(\lambda)$ although, 
technically, the term `convolution'
should be reserved for integrals of the form

$S(l) = \int F_S(\lambda) {\cal A}(\lambda-l)\ d\lambda$,
but note that this is a special case of Eqn.~(\ref{integral1}). 
Suffice it to say, a spectral
line broadened by a spectrometer arises from a convolution and not from a 
product.}
Bland-Hawthorn~(1995) gives a worked example of
standard star flux measurement using a Fabry-Perot and the definitions
above.

\subsection{Overall Instrumental Efficiency}

Efficiency measurements are best made on a nightly basis 
and observed at a range of airmass encompassing the science observations
where possible. 
Good sources of suitable standards are Bessell~(1999) and Hamuy~\etal~(1994),
although the latter have not
removed the effects of atmospheric (telluric) absorption features from
their measurements, with the partially saturated bands 
of H$_2$O subject to variation with humidity and airmass 
(Stevenson~1994). These standards 
are sufficiently bright and isolated
to permit measurement through large
10 -- 20\as\ apertures.
The distance of the star from the optical axis also needs to
be measured and the appropriate wavelength correction 
applied (Bland-Hawthorn \& Jones~1998b).
Since it is impractical to observe a flux standard at 
every pixel position, we can only flux calibrate
the spectral response at each point in the field through the 
whitelight cube. Thus, we effectively calibrate
the whitelight response at the position of the flux standard and 
thereafter the data cube.

The total efficiency $\epsilon(\lambda)$ of the 
telescope, optics and detector
is the ratio $F_{\rm m}(\lambda)/F_{\rm p}(\lambda)$ of measured to published
flux from the standard. Measured fluxes can be converted to true flux 
(\lineunits) from ADU through  
\begin{equation}
F_{\rm m}(\lambda) = \frac{g \cdot K(\lambda) \cdot E_{\gamma}(\lambda)}
{t \cdot A_{\rm tel}} \cdot F_{\rm ADU} (\lambda),
\label{e:effCal}
\end{equation}
where $g$ is the CCD gain (\eADU), $E_{\gamma}(\lambda)$ is
the energy of a photon ($\gamma$) of that wavelength (\ergphot),
$t$ is the exposure
time (s), $A_{\rm tel}$ is the area of the telescope primary (cm$^2$) and 
$\delta \lambda_{\rm e}$ is effective passband width (\AA)
as described in Sect.~\ref{jacquinot}. $K(\lambda)$
is a correction for extinction,
\begin{equation}
K(\lambda)= 10^{0.4 k(\lambda) \langle \chi \rangle},
\label{e:extin}
\end{equation}
dependent upon
both the extinction coefficient $k(\lambda)$ for the site
and the mean airmass $\langle \chi \rangle$ of the observation.

\begin{figure*}
\psfig{file=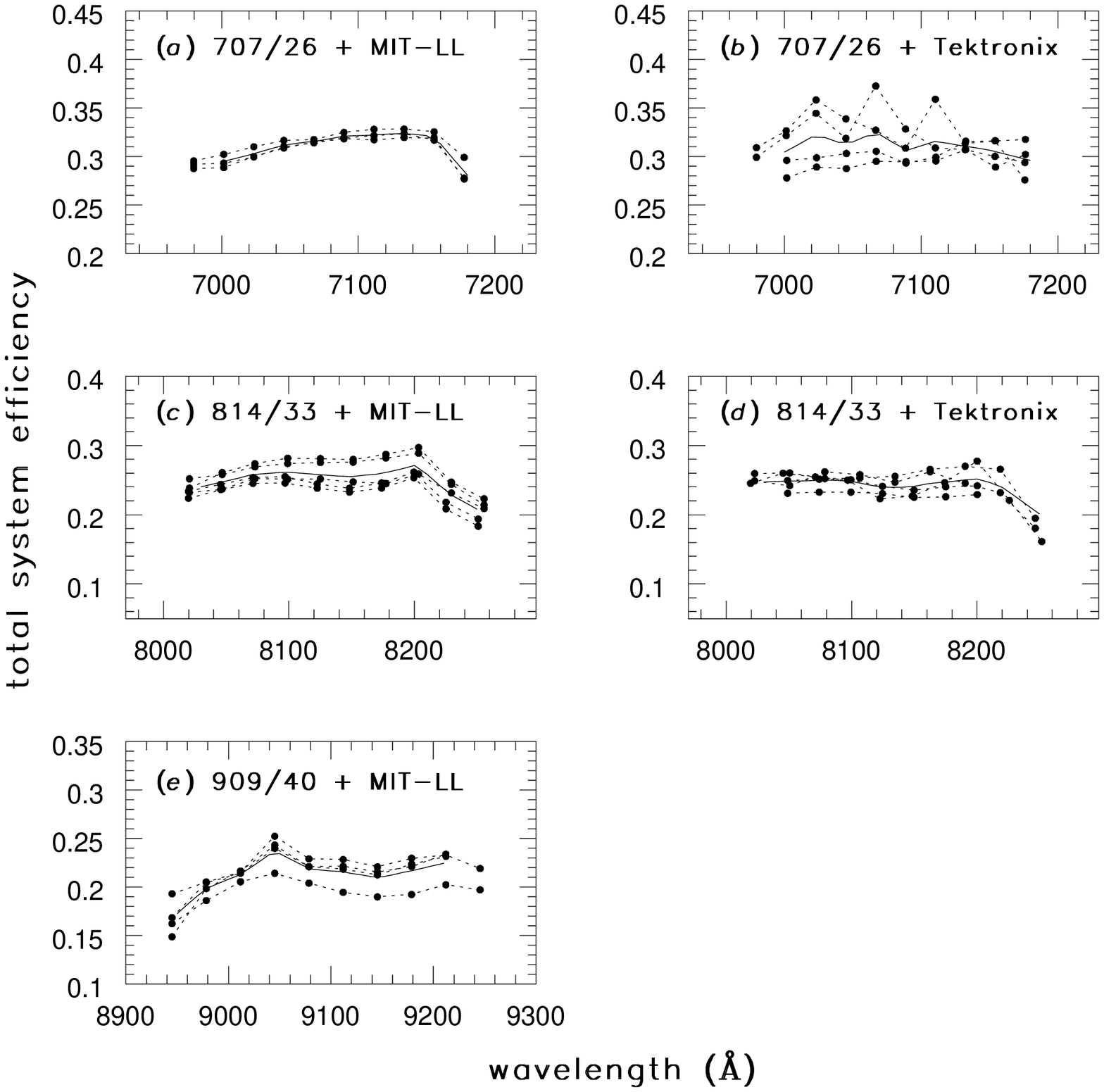,width=0.9\hsize,angle=0}
\caption[Measurements of total system efficiency]
{Measurements of total system efficiency (atmosphere $+$ telescope 
$+$ \taurus2) using both the MIT-LL ($a$,$c$,$e$) and Tektronix
($b$,$d$) CCDs. Scans through the 707/26, 814/33 and 909/40~nm
blocking filters are shown. Individual scans ({\em dotted}) were taken
on observing runs between September 1996 and April 1998. Mean values 
for each combination ({\em solid}) have also been determined.
The drop in efficiency at the edges is due to the spectral cut-off
of the blocking filter.}
\label{f:effic}
\end{figure*}

Fig.~\ref{f:effic} shows nightly efficiency measurements for
the TTF at the AAT with 
two types of CCD (MIT-LL and Tektronix), over observations spread across
four observing semesters (two years).
The scatter about the mean is typically a few percent
in each case. This is most likely due to variations in the photometric
quality between the nights. In this
particular case, the design efficiencies of the individual 
system components are 65\% for \taurus2, 90\% for the AAT, 70\% for 
the CCD and 90\% for TTF. Combined, these yield
$\sim 37$\% throughput, in broad agreement with our measurements.

\subsection{Narrowband Flux Calibration}

With a knowledge of $\epsilon(\lambda)$
we can convert directly from observed flux (in ADU)
to true 
flux (in \ergs2band). In a single observation,
\begin{equation}
f_i(\lambda) = \frac{g \cdot K(\lambda) \cdot E_{\gamma}(\lambda)}
{t \cdot A_{\rm tel} \cdot \epsilon(\lambda)}
 \cdot f_{i,{\rm ADU}} (\lambda).
\label{e:fluxCal}
\end{equation}
This is the {\sl total} flux within the passband, not per unit wavelength. 
Implicit in it is the assumption of a linear CCD detector.

Tunable filter object photometry typically consists of co-added observations 
at different airmass or differing exposure time. If the co-added
flux of an object (in ADU) is the sum of individual frames,
\begin{equation}
F_{\rm ADU}(\lambda) =  \sum_{i=1}^{n} f_{i,{\rm ADU}} (\lambda),
\label{e:fluxSum}
\end{equation}
then that same object will have 
true flux given by 
\begin{equation}
f(\lambda) = F_{\rm ADU}(\lambda) \cdot  \frac{g \cdot E_{\gamma}(\lambda)}
{A_{\rm tel} \cdot \epsilon(\lambda)}   \cdot
\Biggl ( \sum_{i=1}^{n} \frac{t_i}{K_i} \Biggr )^{-1} ,
\label{e:fluxFinal}
\end{equation}
where $g$, $E(\lambda)$ and $\epsilon(\lambda)$ are
constant for exposures of a given wavelength.

At this point, some additional flux calibrations may be necessary
depending on the nature of the data. In point-source 
measurements for which the apertures have been
chosen to optimise signal-to-noise, a correction is
necessary to derive total fluxes. 
One advantage of spectral imaging with a Fabry-Perot is that the object
aperture can be optimised {\sl after} the observation rather than set
at the {\sl time} of observation, as is the case with all slit spectrometers.


\begin{figure} 
\psfig{file=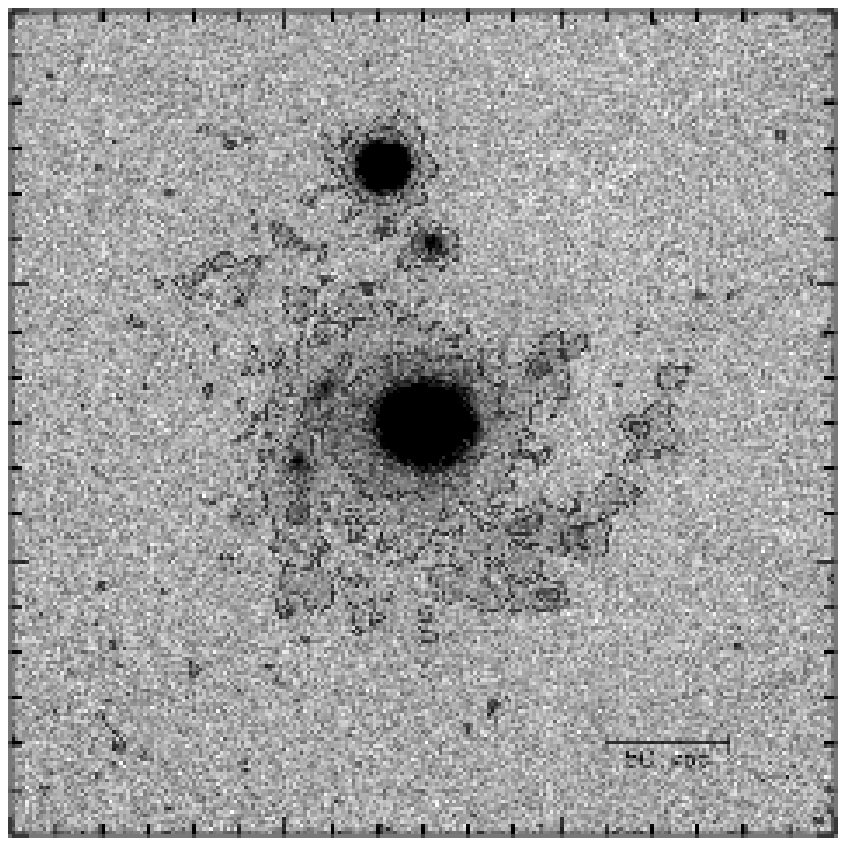,width=\hsize}
\caption{Deep H$\alpha$ image showing the extended nebular emission surrounding
the quasar MR2251-178. The image was obtained by Shopbell, Veilleux \& 
Bland-Hawthorn~(1999) using the Taurus Tunable Filter. 
The frame is $3' \times 2.5'$ with north up and east to 
the left, and reaches a flux limit of  $1.8 \times 10^{-17}$
erg~s$^{-1}$~cm$^{-2}$~arcsec$^{-2}$. The faint curved structure to the 
south-west is real and not a residual ghost image.
}
\label{f:qso}
\end{figure}

\section{SUMMARY AND FUTURE PROSPECTS}

In this review we have outlined the main principles behind Fabry-Perot 
detection and spectrophotometry from narrowband scans of Fabry-Perot
data, with particular reference to low resolution tunable
filter data. We have also discussed the most common limitations 
which need to be addressed for reliable photometric 
calibration. The imaging
Fabry-Perot interferometer has the capability to provide superior 
spectrophotometry, since
slit-aperture devices suffer seeing losses and narrowband filters are 
tacitly assumed to have constant
transmission properties as a function of both position and wavelength. 
Fabry-Perot interferometers are
still not common-user instruments at any observatory for a variety of 
reasons, most notably because
of the restricted wavelength coverage. 
But for studying extended emission from a few bright 
lines, the capabilities of the Fabry-Perot are unmatched by any other 
technique, with the exception of the 
imaging Fourier Transform spectrometer.  
Figure~\ref{f:qso} shows tunable filter imaging of the faint H$\alpha$
extended emission around the $z = 0.0638$ quasar MR2251-178, obtained 
by Shopbell, Veilleux \& Bland-Hawthorn~(1999) using
techniques described in this paper.

Fabry-Perot interferometers (including TTF) differ from an ideal
tunable monochromator in three areas. First, the triangular rather than
square shape of the Airy profile limits sampling efficiency and
complicates flux calibration. A potential solution
is the development of a `double cavity' interferometer which squares
the Lorentzian instrumental profile while maintaining high 
throughput (vand~de~Stadt \& Muller~1985). Second, the use of
fixed blocking filters limits the true tunable potential
of the device. The use of two or more Fabry-Perot interferometers
in series can bypass this restriction, thereby making a fully
tunable system (Meaburn~1972). 
However, such a system comes at the cost of increased operating
complexity. 
An intermediate solution is the use of
custom-built rugate multiband blocking filters, transmitting two
widely-separated intervals simultaneously (Cianci, Bland-Hawthorn \&
O'Byrne~2000).
Finally, the phase change across
the field presents a position-dependent passband that restricts the
angular size of emission that can be imaged at any one time.

However, there are many promising technologies that look
set to take tunable imaging beyond the Fabry-Perot interferometer.
Acousto-optic tunable filters are an alternative where
the active optical element comprises piezoelectric transducers
bonded to an anisotropic birefringent material (Harris \& Wallace~1969).
An AC signal through the transducers (10-250~MHz), creates
sonic waves that vibrate the crystal where it acts like a diffraction
grating. Such tunable
filters are currently limited by the expense and difficulty at 
manufacturing suitably-sized crystals of good imaging quality.
The power to drive the vibrations is high (several watts
per square centimetre) and there are difficulties concerning 
heat dissipation. Liquid crystal filters are an alternative technology
that are now commercially produced, albeit with transmissions
($\sim 30$\%) that remain too low for nighttime astronomical work.
Solc filters (Solc~1959; Evans~1958) are tunable devices utilising 
a pair of polarisers with a sequence of phase retarders set to
differing position angles. With a Solc filter it is possible to tune
the {\sl shape} of the spectral profile, in addition to its
placement and dimensions.

The most advanced alternative tunable technique is that
of the tunable Lyot filter,
(Lyot~1944), consisting of many birefringent 
crystals in series. Birefringent filters cause incident light
of two polarisation states to interfere with each other.
The resulting bandpass is tuned by rotating all elements and its
width is set by the thickness of the thinnest crystal.
Such systems are already in wide use at solar observatories and
we believe it will not be long before such systems find
their way into nighttime work.
The main drawback for astronomical use is that
half the light is lost at the entrance polariser.
The widefield tunable Lyot design of Bland-Hawthorn~\etal~(2001)
places such an arrangement at the prime focus of the telescope.
Not only does this combat the light-loss problem, but it 
gives the filter access to wide sky angles (and short $f$-ratios) 
that would never be feasible with a Fabry-Perot instrument.

In the mean time, however, a low-order Fabry-Perot etalon such as TTF 
presents the best high-throughput system capable of tunable imaging.

\section*{Acknowledgments}

We acknowledge C.~Tinney for the preparation of Fig.~\ref{radial}
and G.~Cecil for Fig.~\ref{f:ghosts}. We also thank
S.~Serjeant who contributed to the 
derivation of the test in Appendix~\ref{weakcosmics}. 
Anglo-Australian Observatory Directors past and present,
R.~D.~Cannon and B.~J.~Boyle, and the Head of Instrumentation,
K.~Taylor, have strongly supported development of 
Fabry-Perot instrumentation at the AAT.
Not least, our thanks are also due to the technical and support staff 
at the AAT for their assistance in implementing new and 
difficult techniques. We would especially like to mention J.~R.~Barton,
E.~J.~Penny, L.~G.~Waller, T.~J.~Farrell and C.~McCowage for all the
necessary upgrades to both hardware and software.
 
We would like to thank J.~Baker and D.~Rupke for their comments on 
earlier versions of this paper, and to the referee J.~Meaburn, whose
comments improved the content of the final draft in several key areas.
Finally, we acknowledge the Australian Time Allocation Committee
(1996--1998) for their generous allocations of AAT time upon which
much of this work was fundamentally dependent.


\section*{References}
\frenchspacing
\begin{refs}
\mnref{Acker, A., Ochensenbein, F., Stenholm, B., Tylenda, R.
Marcout, J., Mischon, C.~1992, The Strasbourg-ESO Catalogue of
Galactic Planetary Nebulae (ESO: Garching)}
\mnref{Atherton, P.~D., Reay, N.~K., Ring, J., Hicks, T.~R.~1981, 
Opt Eng 20, 806}
\mnref{Atherton, P.~D., Reay, N.~K.,~1981, MNRAS 197, 507}
\mnref{Bacon, D.~J., Refregier, A.~R., Ellis, R.~S.~2000, MNRAS 318, 625}
\mnref{Baker, J.~C., Hunstead, R.~W., Bremer, M.~N.,
Bland-Hawthorn, J., Athreya, R.~A.\ and Barr, J.\ 2001, AJ 121, 1821}
\mnref{Bertin, E., Arnouts, S.~1996, A\&A 117, 393}
\mnref{Bessell, M.~S.~1999, PASP 111, 1426}
\mnref{Bland, J., Tully, R.~B.~1989, AJ 98, 723}
\mnref{Bland-Hawthorn, J., Shopbell, P.~L., Cecil, G.~1992, in Astronomical 
Data Analysis: Software \& Systems $-$ I, eds.~D.~Worrall, C.~Biemsderfer,  
J.~Barnes (ASP: San Francisco), p.~393}
\mnref{Bland-Hawthorn, J.~1995, in Tridimensional Optical Spectroscopic 
Methods in Astrophysics, ASP Conference Series Volume 71, eds. G.~Comte,
M.~Marcelin (ASP: San Francisco), p. 72}
\mnref{Bland-Hawthorn, J., Jones, D.~H.~1998a, PASA, 15, 44}
\mnref{Bland-Hawthorn, J., Jones, D.~H.~1998b, in 
Optical Astronomical Instrumentation, Proc~SPIE 3355, 
ed.~S.~D'Odorico, (SPIE: Bellingham, Washington), 855}
\mnref{Bland-Hawthorn, J., van~Breugal, W., Gillingham, P.~R.,
Baldry, I.~K., Jones, D.~H.~2001, ApJ accepted}
\mnref{Bookstein, F.L. 1979, Comp. Graph. \& Image Proc., 9, 56}
\mnref{Cecil, G., 1988, ApJ 329, 38}
\mnref{Cianci, S.~2002, {\sl PhD Dissertation}, University of Sydney, in prep.}
\mnref{Cianci, S., Bland-Hawthorn, J., O'Byrne, J.~W.~2000, in
Optical and IR Telescope Instrumentation and Detectors, Proc~SPIE 4008, 
(SPIE: Bellingham, Washington), 1368}
\mnref{Deutsch, E.W., Margon, B., Bland-Hawthorn, J.~1998, PASP 110, 912}
\mnref{Evans, J.~W..~1958, J Opt Soc Amer 48, 142}
\mnref{Gander, W., von Matt, U.~1993, in Scientific Problems in 
Scientific Computing, eds.~W.~Gander \& J.~Hrebicek, (Springer-Verlag), p~251}
\mnref{Glazebrook, K., Bland-Hawthorn, J., 2001, PASP 113, 197}
\mnref{Haffner, L.\ M., Reynolds, R.\ J., Tufte, S.\ L.\ 1999, ApJ, 523, 223} 
\mnref{Hamuy, M., Suntzeff, N.~B., Heathcote, S.~R., Walker, A.~R., 
Gigoux, P., Phillips, M.~M.~1994, PASP 106, 566}
\mnref{Harris, S.~E., Wallace, R.~W.~1969, J.~Opt.~Soc.~Amer., 59, 744}
\mnref{Jacquinot, P.~1954, J Opt Soc Amer 44, 761}
\mnref{Jacquinot, P.~1960, Rep Prog Phys 23, 267}
\mnref{Jones, D.~H., Bland-Hawthorn, J.~2001, ApJ 550, 593}
\mnref{Jones, D.~H~1999, {\sl PhD Dissertation}, Australian National 
University}
\mnref{Jones, D.~H., Bland-Hawthorn, J.~1998, PASP 110, 1059}
\mnref{Joseph, S.~H.~1994,  Comp Graph \& Image Proc, 56, 424}
\mnref{Laval, A., Boulesteix, J., Georgelin, Y.\ P., Georgelin, Y.\ M., Marcelin, M.\ 1987, A\&A, 175, 199} 
\mnref{Lissberger, P.~H., Wilcock, W.~L.~1959, J Opt Soc Amer 49, 126}
\mnref{Lyot, B.~1944, Ann.~d'Ap., 7, 31}
\mnref{Meaburn, J.~1972, A\&A 17, 106}
\mnref{Meaburn, J.~1976, Detection and Spectrometry of Faint Light, 
Dordrecht: Reidel}
\mnref{Rhoads, J.~E.~2000, PASP 112, 703}
\mnref{Roesler, F.~L.~1974, Meth Expt Phys, 12A, chap. 12}
\mnref{Shopbell, P.~L., Bland-Hawthorn, J.~1998, ApJ 493, 129}
\mnref{Shopbell, P.~L., Veilleux, S., Bland-Hawthorn, J.~1999, ApJ 524, L83}
\mnref{Stilburn, J.~R.~2000, in Optical and IR Telescope Instrumentation and
Detectors, Proc~SPIE 4008, eds.~M.~Iye, A.~F.~Moorwood, 
(SPIE: Bellingham, Washington), 1361}
\mnref{Solc, I.~1959, Czech J Phys, 9, 237}
\mnref{Stevenson, C.~C.~1994, MNRAS 267, 904}
\mnref{Taylor, K., Atherton, P.~D.~1980,  MNRAS 191, 675}
\mnref{Thorne, A.~P.~1988, Spectrophysics, Chapman \& Hall, London}
\mnref{Tinney, C.~G., Tolley, A.~J~1999, MNRAS 304, 119}
\mnref{Valdes, F.~1993, {\sl FOCAS User's Guide}, NOAO}
\mnref{van~de~Stadt, H., Muller, J.~M.~1985, J.~Opt.~Soc.~Am.~A 2, 1363}
\mnref{Veilleux, S., Bland-Hawthorn, J., Cecil, G., ~1999, AJ 118, 2108}

\end{refs}


\begin{appendix}


\section{Fitting Night-Sky Rings}
\label{ringFitting}

{\bf Circular Fitting:} Given three
well-separated positions $(x_1,y_1)$, $(x_2,y_2)$ and $(x_3,y_3)$
along the ring, the centre $(x,y)$ can be found by solving the
determinant
\begin{equation}
\left|
\begin{array}{cccc}
x^2 + y^2 & x  &  y & 1 \\
x_1^2 + y_1^2 & x_1  &  y_1 & 1 \\
x_2^2 + y_2^2 & x_2  &  y_2 & 1 \\
x_3^2 + y_3^2 & x_3  &  y_3 & 1 \\
\end{array}
\right|
= 0 .
\end{equation}
\vspace{5mm}

{\bf Elliptical Fitting:} 
For an ellipse centred on the origin, we write
\begin{equation}
{\bf x}^T A {\bf x} + {\bf b}^T {\bf x} + c = 0
\label{ellip1}
\end{equation}
for which $A$ is symmetric and positive definite.  Here, we have introduced 
a coordinate vector ${\bf x} = (x_{i1},x_{i2})^T$. In order to derive
a form invariant under transformation, we replace ${\bf x}$ with
$G {\bf x} + {\bf h}$. This leads to a similar form to Eqn.~(\ref{ellip1}) 
where $A$ is replaced by $G^T A G$, ${\bf b}$ becomes 
$(2 {\bf h}^T A + {\bf b}^T) G$, 
and  $c$ is replaced by ${\bf h}^T A {\bf h} + {\bf b}^T {\bf h} + c$. 

The eigenvalues $(\lambda_1,\lambda_2)$ for $A$ are obtained from a suitable 
choice of $G$ ($G^T G=1$), and ${\bf h}$ can be chosen to ensure 
${\bf h}^T A {\bf h} + {\bf b}^T {\bf h} + c = 0$. It follows that
\begin{equation}
\lambda_1 {\bf x_1}^2 + \lambda_2 {\bf x_2}^2 + c  = 0
\label{ellip2}
\end{equation}
where $c$ is defined in the transformed frame. This is the equation
of an ellipse subject to $(\lambda_1 > 0, \lambda_2 > 0, c < 0).$
Since $A$ has the same real eigenvalues before and after transformation,
all functions of the eigenvalues are invariant under transformation.

The eigenvalues share an important relationship with the coefficients of $A$,
i.e. 
\begin{eqnarray}
\lambda_1 \lambda_2 &=& A_{11} A_{22} - A_{21} A_{12} \ \ \ ({\rm det~A}) \\
\lambda_1 + \lambda_2 &=& A_{11} +  A_{22} \ \ \ ({\rm trace~A}) \\
\lambda_1^2 + \lambda_2^2 &=& A_{11}^2 + 2 A_{12} + A_{22}^2
\label{ellip3}
\end{eqnarray}
Moreover, the geometric properties of the ellipse before transformation
are
\begin{eqnarray}
\alpha &=& \sqrt{-c/\lambda_1} \\
\beta &=& \sqrt{-c/\lambda_2}
\label{ellip4}
\end{eqnarray}
where $(\alpha,\beta)$ are the semi-major axes; ${\bf h}$ defines
the origin of the ellipse.

Finally, once each of the $p$ data points are placed in Eqn.~(\ref{ellip1}),
we arrive at a highly overdetermined, nonlinear system of $p$ equations 
in the 4 unknown geometric parameters.  This is a standard problem which 
can be solved efficiently with any number of gradient search algorithms
subject to a constraint. The restriction needs to be invariant to the
coordinate transformation: simply setting one of the unknowns to unity
is not normally sufficient to ensure convergence. Two useful constraints
(everywhere non-zero) are $\lambda_1+\lambda_2=1$ (Gander \& von~Matt~1993) 
and $\lambda_1^2+\lambda_2^2=1$ (Bookstein~1979).

It is important that the method, e.g. Levenberg-Marquardt, be efficient
since $p \sim 10^3$ typically in TTF analysis.  There are several possible 
weighting schemes which have varying merits for different data sets. The 
most common is flux weighting to ensure the fitting establishes the major 
axes (radius) as accurately as possible.


\section{Ghost Families}
\label{ghostfamilies}

Internal reflections constitute a challenge 
for all optical systems, particularly
imagers with dispersing optics at the pupil,
such as tunable filters and Wollaston prisms.
Even a minimal arrangement can have eight or 
more optically flat surfaces ({\em e.g.}~blocking
filter, CCD window).  At some level, all of these 
surfaces interact separately to generate 
spurious reflections.

\begin{figure}
\psfig{file=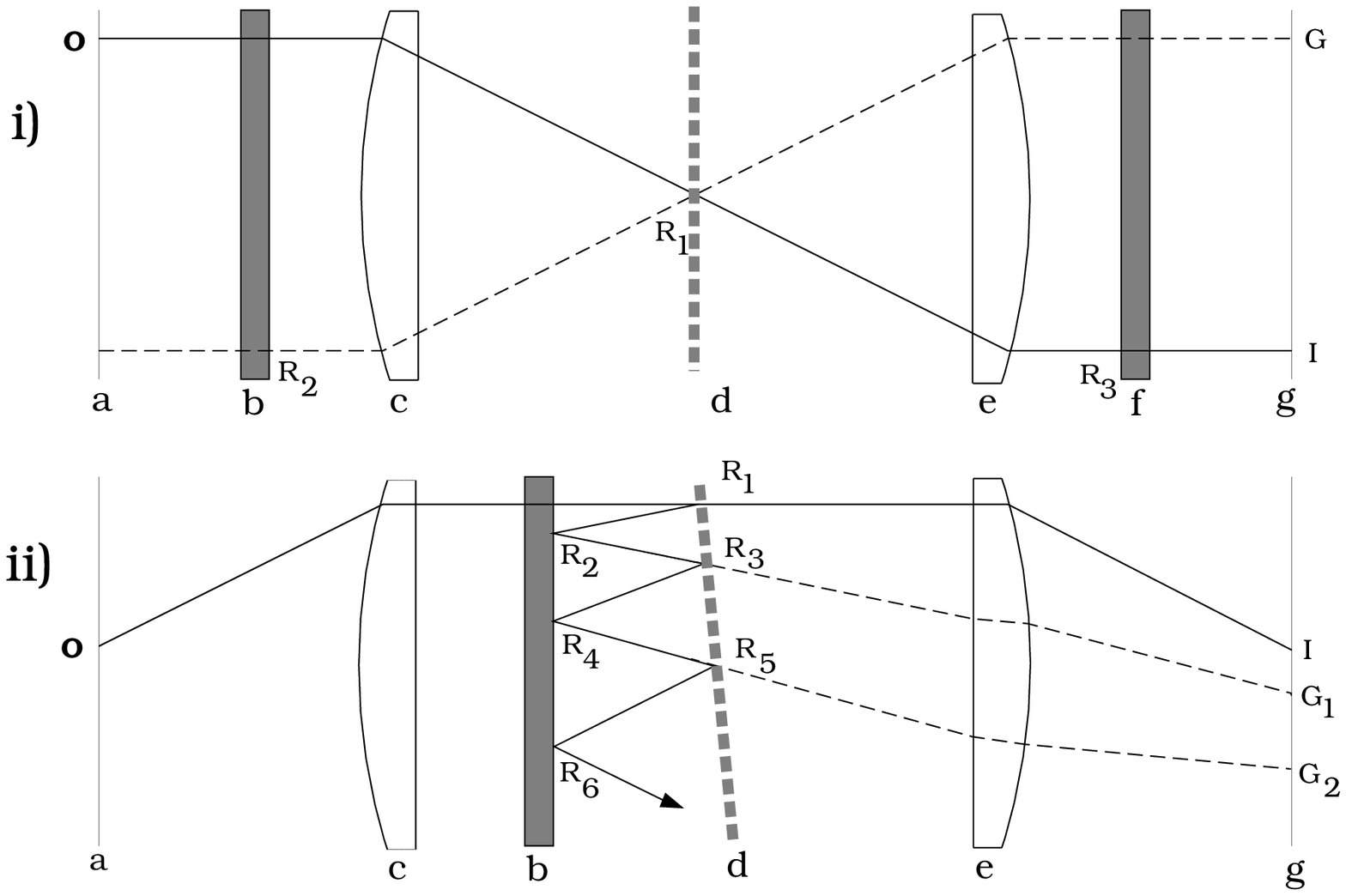,width=\hsize}
\caption{Ghost families arising from internal reflections within a 
Fabry-Perot spectrometer. ($i$) Diametric ghosts. Rays from the object O 
form an inverted image I
and an out-of-focus image at R$_3$. The reflection at R$_1$ produces 
an out-of-focus
image at R$_2$. The images at R$_2$ and R$_3$ appear as a ghost image 
G at the
detector.  ($ii$) Exponential ghosts. The images at R$_2$ and R$_4$ 
appear as ghost
images G$_1$ and G$_2$ respectively.}
\label{f:rays}
\end{figure}

\begin{figure}
\psfig{file=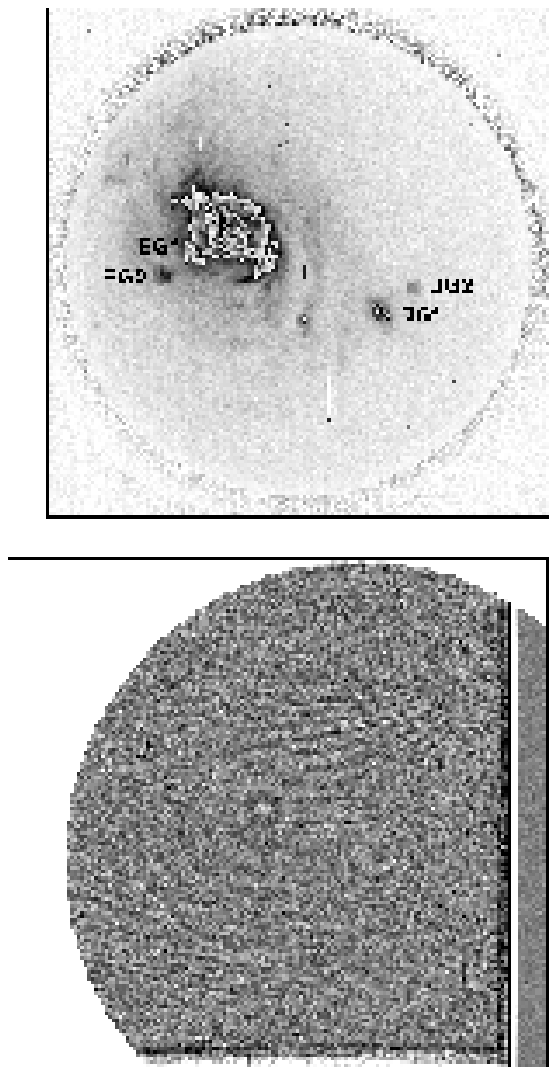,width=\hsize}
\caption{{\em Top.}- Two ghost families are seen in this image 
of NGC~1068 taken with the 
Rutgers Fabry-Perot on the CTIO 4.0~m telescope. The optical axis is 
indicated by the cross.
{\bf N} is placed slightly to the north of the Seyfert nucleus. {\bf DG1} is 
the diametric ghost of the active 
nucleus; {\bf EG1} is an exponential ghost of the Seyfert nucleus. {\bf DG2} 
is the diametric
ghost of {\bf EG1} and {\bf EG2} is the exponential ghost of {\bf EG1}.
{\em Bottom.-} The `extraneous etalon' ghost pattern from the downstream 
etalon plate (lower right 
quadrant) and from an air gap in the MOSFP camera (central). The data 
were taken at the CFHT 
by illuminating the dome with an H$\alpha$ lamp. The fringe pattern 
of the Loral \#3 CCD has a 
similar peak to trough amplitude.}
\label{f:ghosts}
\end{figure}

The periodic behaviour of the tunable filter always 
requires that we use an interference filter 
somewhere in the optical path, to block light from extraneous
orders. The filter introduces ghost 
reflections within the Fabry-Perot optics. 
Such a filter placed 
in the converging beam before
the collimator or after the camera lens generates 
a distinct pattern of {\sl diametric} ghosts.
A filter placed in the collimated beam has 
its own distinct family of {\sl exponential} 
ghosts which are the most difficult to combat. 
The pattern of ghosts imaged at the detector is
different in both arrangements, as illustrated in 
Fig.~\ref{f:rays}. Examples of the appearance of
such ghosts are shown in Fig.~\ref{f:ghosts}.

Another ghost family arises from the down-stream 
optical element ({\sl extraneous etalon} effect).
The large optical gap of the outer plate surfaces produces
a high-order Airy pattern at the detector.
It is possible to suppress this signal by wedging 
the outer surfaces.
However, if we are forced to operate the blocking 
filters in the collimated 
beam, this compounds the problem further.

To a large extent, we can minimize the impact 
of diametric and exponential ghosts through 
the use of well-designed anti-reflection coatings. 
The reflectivity $r$ of an air-glass interface is
\begin{equation}
r = \frac {(\mu_{\rm glass} - \mu_{\rm air})^2} {(\mu_{\rm glass} + 
\mu_{air})^2} = 4\%
\end{equation}
where $\mu_{\rm glass} \approx 1.5$ and $\mu_{\rm air}=1.0$.
This can be significantly reduced to 1\% or better 
by the application of an AR 
coating (half-optical band), or to below 0.1\% for 
AR coatings customized to the particular
filter in question. Coating performance
is currently limited by the availability of pure 
transparent dielectrics with high refractive indices.
A promising alternative is the use of silica sol-gel
coatings, that reduce the reflective losses inherent with 
traditional MgF$_2$ coatings (Stilburn~2000). The durability
of these coatings has been improved over recent years to the
point where they are now being routinely used on instruments
for 4 and 8~m-class telescopes with many air-glass surfaces.

\begin{figure}
\psfig{file=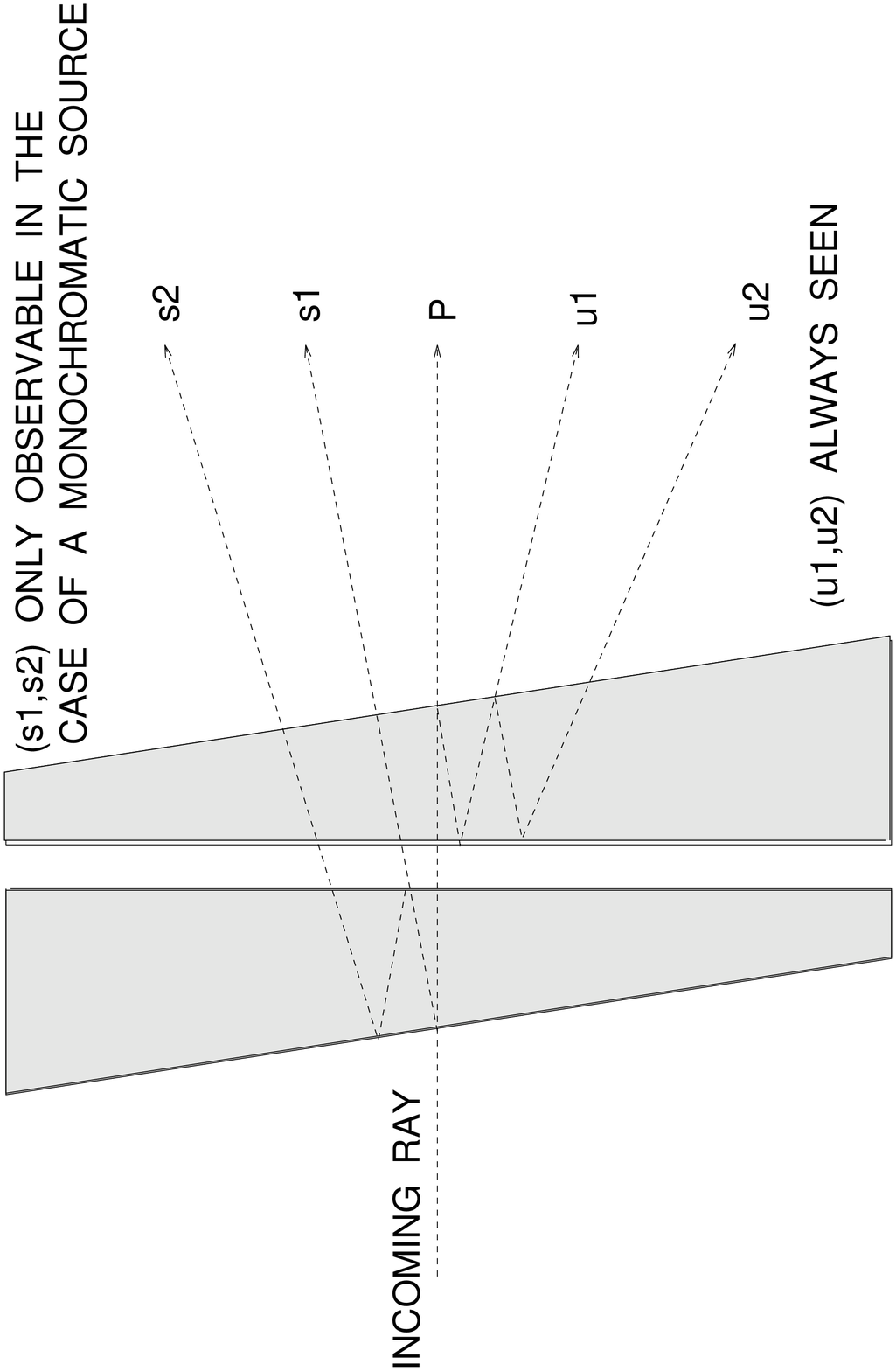,width=0.9\hsize,angle=270}
\caption{The diverging ghost families arising from wedged Fabry-Perot
plates with poor anti-reflection coatings on the outer surfaces.}
\label{wedge_ghosts}
\end{figure}

A more difficult reflection problem arises from the 
optical blanks which form 
the basis of the etalon. These can 
act as internally reflecting cavities since, from Sect.~\ref{response}, if 
we let ${\cal R}_1 = 0.96$ and ${\cal R}_2 = {\cal AR}_1 = 0.04$ 
(air-glass), it generates a ripple pattern with a 
finesse close to unity. The large optical gap of 
the outer plates produces a high-order Airy pattern at the 
detector (Fig.~\ref{f:ghosts}). Traditionally, the outer surfaces have
been wedge-shaped to deflect this spurious signal out of the beam. 
However, if the surfaces are not coated with suitable
anti-reflection coatings then the problem is compounded by
an additional set of ghosts (Fig.~\ref{wedge_ghosts}). These
spurious reflections diverge as a result of the wedge angle.
Furthermore, reflections from the front plate only occur for
monochromatic sources, thereby complicating identification
in fields containing both stars and emission-line sources.

Even after we have paid full attention to 
minimizing {\sl air-glass} reflections, there 
remains one fundamental ghost arising from the CCD.  
As we see from the equation above,
silicon ($\mu_{\rm Si}= 3.5$) leads to an 
air-silicon reflection of 30\%. A modern prescription 
broadband AR coating for a CCD reduces this by a factor of 3, 
meaning that ghosts reside at around 10\%.  Since we
cannot customize the CCD response for a narrow wavelength 
range, bright `in-focus' diametric
ghosts must always occur with a tunable filter imager.
A good way to track these down
is to place a regular grid of holes in focus at the focal plane and 
illuminate the optical system 
with a whitelight source (Sect.~\ref{holes}), tilting the etalon
such that ghost images of the grid avoid the detector area.
If tilting is not possible, then dithering and median filtering are
required, (Sect.~\ref{s:cosmicRay}).

\section{Weak Cosmic Rays}
\label{weakcosmics}

Here, we describe a simple test to evaluate the effectiveness 
of any software in removing weak cosmic-ray events. The method is 
to compare the probability distribution function (PDF) 
of cosmic ray events between an observation and a dark frame matched in 
exposure time. We illustrate the basic idea using dark exposures from a
Tektronix $1024 \times 1024$ CCD with varying exposure lengths and 
read-out times.

We used the Tekronix CCD at the
AAT 3.9~m with  varying exposure lengths (15, 30, 60, 120~min) and read-out 
times (FAST, SLOW, XTRASLOW). The histogram 
of each frame shows the contribution from the bias, read and dark noise. A
millisecond exposure was used to remove the bias and read noise contribution
to each histogram. The additional contribution from the dark noise was well 
calibrated at 0.11 cts pix$^{-1}$ ksec$^{-1}$. It is assumed that the 
remaining events are related to cosmic rays. 

\begin{figure}
\psfig{file=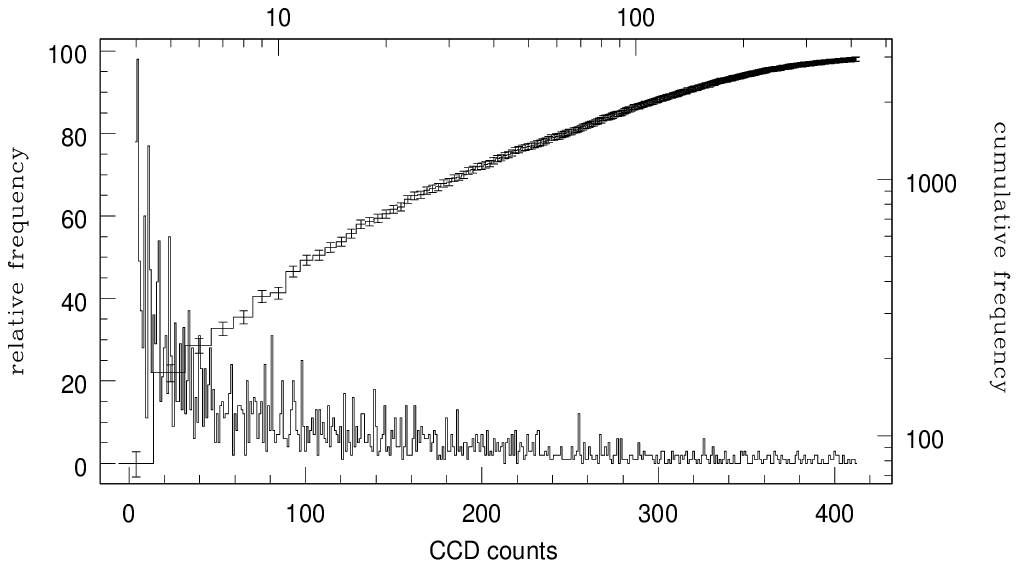,width=0.9\hsize}
\vspace{0.5cm}
\psfig{file=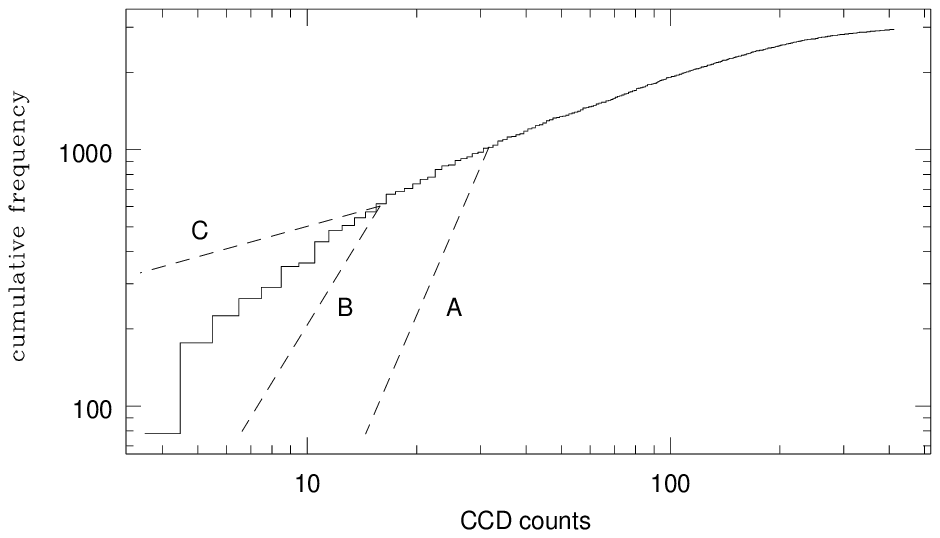,width=0.9\hsize}
\caption{
{\em Top.} Histogram of cosmic ray events in a 
two hour dark frame. The monotonic
curve is the cumulative histogram of these events. The error bars
are Poissonian and not independent.
{\em Bottom.} The cumulative histogram from {\em above} with which to 
compare the performance
of a deglitch algorithm. Three cases are illustrated: algorithm A
is too conservative, algorithm B more reliable, and algorithm C has mistaken
real data with faint cosmic rays.}
\label{f:cosmics}
\end{figure}

We define $n(E)\ dE$ to be the number of cosmic ray events with energies 
(expressed in counts) in the range $(E,E+dE)$. The cumulative distribution 
is then
\begin{equation}
P(E) = \int_{0}^{E} n(E)\ dE
\end{equation}
When $n(E) \propto E^{\beta}$, the slope
of the plot $\log P$ vs. $\log E$ is proportional to $\beta + 1$. In 
Fig.~\ref{f:cosmics}~({\em top}),
the noisy histogram is the bias/dark/read noise subtracted dark frame. The 
monotonic curve is a plot of $\log P$ vs. $\log E$ which is found to be rather 
well defined and reproducible over the different exposures. The PDF
for the data is determined from all events identified by the deglitch
algorithm. Since the energetic events are easier to find, the bright end
of both PDFs will be well matched. In Fig.~\ref{f:cosmics}~({\em bottom}), 
where the deglitch PDF turns over at low energy 
--- presumably but not necessarily at an energy greater 
than or equal to the turnover in the dark PDF --- gives some idea as to how 
effective the algorithm has been in removing the weaker events.

\end{appendix}

\newpage
\onecolumn

\begin{table}
\centering
\begin{tabular}{lll}
\hline
&&\\
    & $f(x)$ & $\int f(x)\ dx$ \\
&&\\ \hline
&&\\
G & $\exp[-\ln 16\ {\rm mod}[x,\Delta x]^2 / (\delta x)^2]$ & 
$({{\delta x}\over 2}) \sqrt{{\pi\over{\ln 2}}}\; 
{\rm erf}[({{2\sqrt{\ln 2}}\over{\delta x}})\ {\rm mod}[x,\Delta x]]$ \\
&&\\
L & $(1+({2\over {\delta x}})^2 \ {\rm mod}[x,\Delta x]^2)^{-1}$ & 
$({{\delta x}\over 2})\ \tan^{-1} [({2\over{\delta x}})\ {\rm mod}[x,\Delta x]]$ \\
&&\\  
A & $(1+ \alpha \sin^2({{\pi x}\over{\Delta x}}))^{-1}$ & 
$({{\Delta x}\over{\pi \sqrt{1+\alpha}}})\ \tan^{-1} [ \sqrt{1+\alpha}\ \tan ({{\pi x}\over{\Delta x}})]$ \\
&&\\ \hline
\end{tabular}
\caption{Cyclic functions which are periodic over $\Delta x$ with FWHM 
$\delta x$.
The (G)aussian, (L)orentzian and (A)iry functions are illustrated in 
Fig.~\ref{f:cyclic}. The {\it mod} function 
is the modulo function and $\alpha = ({{2\ \Delta x}\over{\pi\ \delta x}})^2$. 
Note that for large $\alpha$,
$({{\Delta x}\over{\pi \sqrt{1+\alpha}}}) \approx ({{\delta x}\over 2})$. 
The gap scanning variable $x$ is offset by
${{\Delta x}\over{2}}$ in practice.}
\label{t:cyclicfunctions}
\end{table}

\begin{table}
\centering
\begin{tabular}{cc}
\hline
	&     	\\
 Source & Useful Range\\
        &  (nm)    \\
	&     	\\
\hline
	&     	\\
 zinc	& 308 -- 636  \\
 thallium & 352 -- 535    	\\
 mercury-cadmium  & 370 -- 450   \\
 deuterium-helium & 450 -- 550$^\dagger$   \\
 cesium & 456 -- 921  \\
 neon		& 500$^\dagger$ -- 700   \\
 copper-argon	& 700 -- 1000   \\
	&     	\\
\hline
\end{tabular}
\caption{Arc lamps used for wavelength calibration of the tunable filter
system at the AAT, showing the range of wavelengths over which they
are useful. $\dagger$ For calibrations around 500~nm, the
astrophysical lines \Hb\ and \Oiii\ found in planetary nebulae
(Acker~\etal~1992) may be more suitable.}
\label{t:lamps}
\end{table}

\end{document}